# Weiss-Weinstein bound of frequency estimation error for very weak GNSS signals

Xin Zhang, Xingqun Zhan, Jihong Huang, Jiahui Liu, and Yingchao Xiao
School of Aeronautics and Astronautics, Shanghai Jiao Tong University, Shanghai, China

**Correspondence**
*Xin Zhang*
Room A332, School of Aeronautics and Astronautics, Shanghai Jiao Tong University,
Dong Chuan Road 800, Shanghai, China.
E-mail: xin.zhang@sjtu.edu.cn

**Abstract**

Tightness remains the center quest in all modern estimation bounds. For very weak signals, this is made possible with judicial choices of prior probability distribution and bound family. While current bounds in GNSS assess performance of carrier frequency estimators under Gaussian or uniform assumptions, the circular nature of frequency is overlooked. In addition, of all bounds in Bayesian framework, Weiss-Weinstein bound (WWB) stands out since it is free from regularity conditions or requirements on the prior distribution. Therefore, WWB is extended for the current frequency estimation problem. A divide-and-conquer type of hyperparameter tuning method is developed to level off the curse of computational complexity for the WWB family while enhancing tightness. Synthetic results show that with von Mises as prior probability distribution, WWB provides a bound up to 22.5% tighter than Ziv-Zakaï bound (ZZB) when SNR varies between -3.5 dB and -20 dB, where GNSS signal is deemed extremely weak.

## 1. INTRODUCTION

The primary objective in a navigation system is to estimate a parameter vector $\boldsymbol{\theta}$. In the first category of problems, the parameters are assumed constant or *static* during the observation interval. There are in turn, two subcategories. The first one models the static parameter as a set of *deterministic* or nonrandom yet unknown variables while the second models the parameter set as random with an *a priori* probability distribution function (pdf). Within the collection of deterministic bounds, there are *local* bounds including Cramér Rao bound (CRB) (p480, Cramér, 1949) and Bhattacharyya bound (Bhattacharyya, 1946) which describes the smallest achievable estimation error variance for maximum likelihood (ML) estimators. Since these two local bounds are good performance indicators only in large-sample or high signal-to-noise ratio (SNR) settings, they are often called *asymptotic* or small-error bounds since if we plot them against SNR, they are *tight* in the high SNR region in the sense that the estimated variance approaches but never reaches the true variance. The difference between CRB and Bhattacharyya bound lies in the choice of *score function* (Richmond & Horowitz, 2015), which is a function of $\theta$ to describe the undulation of its log-likelihood function $\phi$: CRB uses first derivative of $\phi$ with respect to the *true* parameter, $\theta$, whereas Bhattacharyya bound extends to use a mixture of unlimited higher order derivatives. The latter choice of score functions help to improve the bound's ability to predict error covariance in the asymptotic region but, it is, however, this very nature of characterizing undulation of log-likelihood function around the neighborhood of the true parameter that fails these local bounds in low SNR regions. The SNR in these cases will be so low that there are no obvious peaks in the log-likelihood function and it is highly probable that wrong peaks which do not correspond to the ML estimate are picked and estimation error is thus wrongly reported as low.

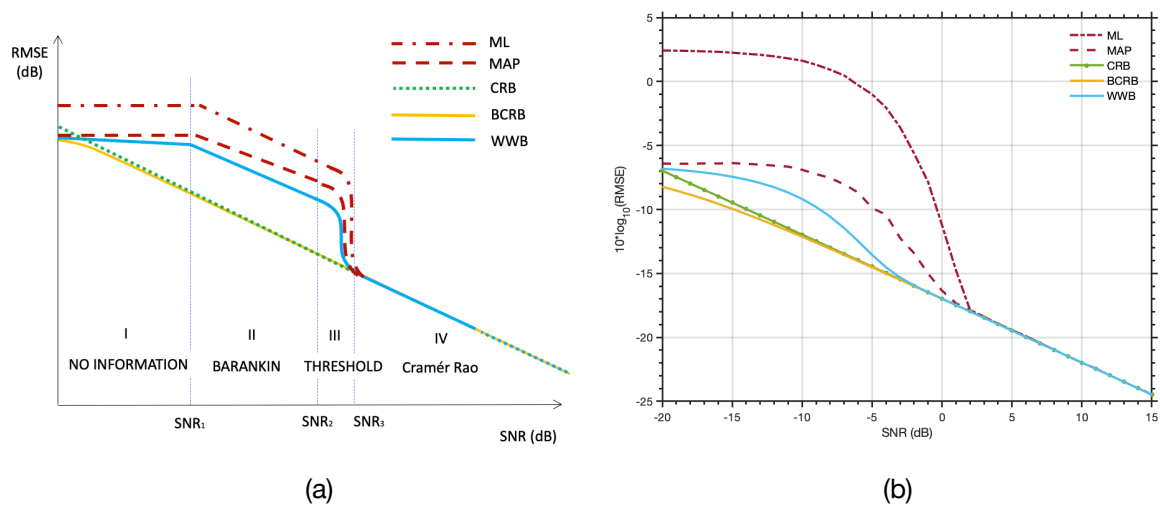

(a)            (b)

**FIGURE 1** Thresholding behavior of root MSE (RMSE) for ML & MAP estimators when SNR transits from medium to low values. (a) An illustrative and delimited bound model by Weiss & Weinstein with CRB, BCRB, and WWB. (b) A simulation of GNSS L1 C/A frequency estimation (tracking) with frequency pull-in range of $[-500, 500]$ Hz using the same bounds and estimators from (a); von Mises, a circular analog of Gaussian distribution is assumed as *a priori* distribution for BCRB and WWB. Here the mean and concentration parameter of von Mises is zero and 20, respectively.

In this regard, a very natural remedy is to include measures of *test points* distinctively different from the true value, $\theta$, in the score function. Test points are estimated parameters supposed to be 'outliers' that lead to the largest possible mean square error (MSE). The first of this kind is due to Barankin (1949) who in turn was inspired by the work of Riesz in functional analysis. A more engineer-friendly derivation of Barankin bound is provided by McAulay and Hofstetter (1971). Barankin bound endorses a much tighter bound in low-to-medium SNR regions and thus can be seen as a large-error or global bound in stark contrast to CRB. We adopt a delimited approach to describing distinctive bound behaviors over the entire SNR range from Weiss & Weinstein's work (1983, Figure 3) and paraphrase it in FIGURE 1 (a) where the delimiters can only be defined qualitatively. The involved bounds are materialized as CRB, and two bounds belonging to the next subcategory: Weiss Weinstein bound (WWB) and Bayesian CRB (BCRB), whose definitions are deferred to Sections 3. In FIGURE 1(b), a simulation of ML and maximum *a posteriori* (MAP) frequency estimators with 10,000 trials, CRB, BCRB, and WWB is shown. One can immediately observe that CRB tends to underestimate MSE while Barankin bound (region II) tends to partially overcome this shortcoming of CRB. However, there are two things left to desire. Firstly, when SNR gets even lower, i.e., in region I of FIGURE 1(a), Barankin bound is unable to characterize correctly the error variance. Secondly, even in the medium-to-low SNR region, the vanilla Barankin bound is still not tight enough (regions II & III of FIGURE 1(a)). For the first problem, the observation that in this area, noise instead of useful signals dominates the observed samples endorses modelling the estimated variable as random, i.e. with a proper *a priori* pdf. Since now we are actually scoring undulation of an *a posteriori function* and developing a bound for MAP estimators, this subcategory of static performance bounds is called Bayesian bounds, which can be divided into two classes: the Ziv-Zakaï (ZZB) family and the Weiss-Weinstein (WWB) family. For the second problem, the most efficient solution is to select test points wisely; this is the central topic of this work and will be revealed in the successive sections.

Unlike all the other bounds we have mentioned so far which are derived from covariance inequalities, the ZZB family is derived from the probability of error in a binary hypothesis testing problem. It includes the original form developed by Ziv and Zakaï (1969), and later developments by Seidman (1970), Chazan et al., (1975), Bellini & Tartara (1974, 1975), Weinstein (1988), and Brown & Liu (1993). The most important development along the line is the extended Ziv-Zakaï bound (EZZB) by Bell et al. (1997), which extended the bound to include vector parameters with arbitrary *a priori* distributions while its predecessors can only accept a scalar parameter with a uniform *a priori* pdf.

In contrast, the WWB family is derived from Cauchy-Schwarz inequality (Ibragimov & Hasminskii, 1981; Lehmann, 1983). Bobrovsky & Zakaï (1975) developed the Bayesian version of Barankin bound as a prelude. Weiss and Weinstein (Weiss & Weinstein, 1985; Weinstein & Weiss, 1988) developed the vanilla WWB. Other members of this family include Bayesian Bhattacharyya bound (Van Trees, 1968, pp.148, problem 2.4.23), Bobrovsky-Zakaï bound (BZB) (Bobrovsky & Zakaï, 1976), and Bayesian Abel lower bound (BAB) (Renaux et al., 2008). Renaux et al. (2008) also proposed a constrained optimization framework to uncover the relationship between Bayesian CRB (BCRB), BZB, BAB, and vanilla WWB in the WWB family where the proposed constraints provide an enlightening view on how to choose the test points. A more recent treatment on WWB can be found in (Chaumette et al., 2017, 2018). As a sidenote, the second problem of interest in navigation is the estimation of a *sample function* of a stochastic process which can be modelled in either a correlation approach or a state dynamics approach where the recursive version of BZB has been developed by Fritsche et al. (2018) and recursive WWB developed by Reece et al. (2005) and Rapoport & Oshman (2004, 2007a, 2007b).

Both WWB and ZZB have long been applied to digital communications systems (Villares, 2005) and positioning (Graff et al., 2021) theoretically and practically (Manlaney 2020; Watson et al., 2020). In terms of GNSS, Denis (2009, chapter 5&6) investigated bounding GNSS code tracking with ZZB. Closas (2009, chapter 5) compared BCRB with vanilla CRB when bounding and comparing errors of 2-step positioning and direct positioning. Later, Gusi-Amigó (2018) adapted EZZB to bound direct positioning error. Gifford et al. (2022) investigates time-of-arrival (TOA) error bounds for wideband and ultrawide band (UWB) ranging systems provided with different *a priori* information on the multipath phenomena.

Finally, the problem of bounding carrier tracking and in particular, frequency tracking using modern and theoretically tight bounds such as WWB and comparison between WWB and ZZB has been overlooked, although code tracking using ZZB under Gaussian assumptions *has* been investigated (Emmanuele, 2012). There *are* asymptotic derivations for frequency estimation, for the standard case, high dynamic case (Mcphee et al., 2023a) which accounts for up to 100g acceleration long the line-of-sight (LOS), and a generic non-line-of-sight (NLOS) case (Lubeigt et al., 2020). The misspecified CRB (Ortega, 2023; McPhee, 2023b; Lubeigt et al., 2023) have been thoroughly investigated – misspecified because the assumed signal model unintentionally and practically differs from the received signal, resulting in bias. Therefore, the misspecified CRB describes the performance of biased estimators at low SNR, where RMSE is characterized by bias (which may be large for many cases) combined with the misspecified CRB. However, because the ambiguity function of the Doppler parameter is much wider than that of the delay, the convergence of the Doppler estimate is much earlier than the operating range of the navigation system, e.g. for GPS between 2-3 dB earlier, providing a theoretical limit of frequency estimation at low SNR and hence, for new applications such as Doppler positioning using low earth orbit (LEO) satellites, where the $CN_0$ could be as low as 10 dB-Hz, these asymptotic bounds are not fit enough to fully recover the threshold, Barankin, and no-information regions defined in FIGURE 1(a). Renaux (2008) investigated bounding frequency using WWB, yet with simplified signal model and overlooking the effects of pseudorandom code and assumed a Gaussian *a priori* pdf. All these inspire us to develop a tight bound and in particular, WWB for frequency tracking. The reason for choosing WWB is three-fold: (a) with its test points finetuned, WWB is expected to outperform ZZB in some situations, and (b) WWB is essentially free of regularity conditions (Weiss & Weinstein, 1988), which are haunting other bounds and it is in this regard, WWB is expected to be applied to a broader range of estimation problems.

In this work, we develop WWB for a specific GNSS problem, i.e. frequency tracking and compare it to the other state-of-the-art bound and in particular ZZB for very weak GNSS signals, to clarify the preference of these two bounds. A modern professional GNSS receiver can track signals at a carrier-to-noise-density ratio ($C/N_0$) down to 10 dB-Hz (Manzano-Jurado et al., 2014; Ren & Petovello, 2017; Pany & Eissfeller, 2006) and a typical lower bound of 15 dB-Hz (Soloviev et al., 2009) is required in space applications such as Space Service Volume (IS-GPS-200K, 2019, section 3.3.1.6.1). There exists modern mass-market receiver capable of steady tracking at a received power as low as -197 dBW (U-blox, 2018) which, for a typical front-end with a noise density between -203 and -205 dBW/Hz, means the working $C/N_0$ is as low as 6 to 8 dB-Hz. However, tracking at this level of $C/N_0$ can be extremely challenging, almost impossible. Therefore, input $C/N_0$ between 10 and 45 dB-Hz is assumed in our analysis to cover various GNSS applications.

The remaining sections are organized as follows. Firstly, the estimation problem is formulated using the signal and noise models introduced in section 2 where the *a priori* distribution, von Mises distribution is briefly discussed. The development of the Weiss

Weinstein bound follows in section 3 and the analytical form of GNSS frequency WWB is presented. Detailed development is provided in Appendix A. In section 4, this WWB is optimized over its two model hyperparameters, namely the exponential index $\{s_i\}$ and the test points $\{h_i\}$ where $i = 1, \ldots, R$; $R$ is the number of test points. A divide and conquer optimization procedure including only two steps is introduced to ease optimization over a possibly large sets of parameters $\{s_i, h_i\}$. In particular, a new set of test points is proposed by combining side lobe positions of the estimates and a vector of points evenly spanning the frequency uncertainty region. This optimization can offer a performance advantage of up to 0.95 dB, equivalent to a 19.8 % increase in predicted root mean square error (RMSE), compared to the WWB evaluated using legacy test points. Together with the synthetic simulations in section 5, we demonstrate important properties of the WWB, namely, that a.) with *a priori* pdf of von Mises distribution, WWB is a tighter bound than ZZB in the SNR region of -3.5 ~ -20 dB, where GNSS signal is deemed weak, but inferior to ZZB in predicting the exact thresholding SNR, and b.) WWB and ZZB converge both asymptotically (in high SNR) and in the no-information region. We also investigate the dependence of the developed WWB on *a priori* pdf, input SNR and integration time and found that when von Mises distribution is concerned, the effects of the location parameter, $\mu$ and the concentration parameter, $\kappa$, of WWB, can be decoupled within a limited range. In summary, contributions of this paper are expected to be two-fold:

(1) Better prediction of RMSE of any circular frequency estimator in the mid-to-low transition region of SNR, by developing a frequency WWB using von Mises distribution as prior, and

(2) A divide-and-conquer type of parameter tuning method to choose test points of WWB to achieve a tighter bound.

## 2. PROBLEM FORMULATION

Firstly, we introduce notations including the signal and noise models. Then, we formulate the problem as developing a tight bound on estimation error of the normalized digital frequency. We also confirm that very weak GNSS signals do occur within the transition and even lower SNR region that WWB is trying to characterize. Finally, we introduce von Mises as the chosen prior distribution.

### 2.1 Notations

We will use bold and lower-case letters to denote vectors, bold and upper-case letters to denote matrix, and italic letters to denote scalars. $E_{\mathbf{x},\boldsymbol{\theta}}\{\cdot\}$ means expectation of the curly bracketed function with respect to joint pdf of $\mathbf{x}$ and $\boldsymbol{\theta}$. For an observation vector $\mathbf{x}$ of length $K$ and parameter vector $\boldsymbol{\theta}$ of length $D$, we will denote *a priori* pdf as $p(\boldsymbol{\theta})$, joint pdf of $\mathbf{x}$ and $\boldsymbol{\theta}$ as $p(\mathbf{x}, \boldsymbol{\theta})$ and likelihood function as $p(\mathbf{x}; \boldsymbol{\theta})$ or $p(\mathbf{x}|\boldsymbol{\theta})$ and *a posteriori* pdf as $p(\boldsymbol{\theta}|\mathbf{x})$.

## 2.2 Signal and Noise Models

We consider receiving a complex analytical signal (Betz, 2015)

$$s(t) = \sqrt{C}\gamma(t-\tau)\exp\{i2\pi f_{RF}(t-\tau)\} \tag{1}$$

transmitted over a carrier frequency $f_{RF}$ with delay $\tau$. $C$ is the power in the data/pilot component and $\gamma$ is the baseband signal with bandwidth $B$ including both spreading code and navigation data. This signal model implicitly implies a narrowband assumption on $\gamma$ which means the effect of Doppler dilatation on the baseband signal (Lubeigt, 2020) can be neglected. Mixing in the receiver frontend, using an estimate of the received carrier frequency $\hat{f}_{RF}$, performs frequency translation and adjustment of $s(t)$ as

$$\sqrt{C}\gamma(t-\tau)\exp\{i2\pi f_{RF}(t-\tau)\}\exp\{-i2\pi \hat{f}_{RF}t\} = \sqrt{C}\gamma(t-\tau)\exp\{i2\pi(f_{IF}t - f_{IF}\tau)\} \tag{2}$$

where the residual or the intermediate frequency (IF) after downconversion is $f_{IF} \triangleq f_{RF} - \hat{f}_{RF}$. Frequency tracking then estimates the residual frequency $f_{IF}$. It is assumed that the receiver is approximately synchronized to the overlaid baseband waveform and carrier so correlation may be done by integration and dump at an interval of $1/f_{INT}$. The correlation samples, denoted as $x_k$, can be summarized as (Betz, 2015, eqn. (18.6), (18.8), or (18.10))

$$x_k = r_k + n_k \qquad k = 0,1,\ldots,K-1 \tag{3}$$

where each sample comprises a deterministic part, $r_k$, defined as

$$r_k \triangleq Ae^{i(2\pi f_{IF}t_k + \varphi)} \tag{4}$$

with time epochs $t_k \triangleq k/f_{INT}$ and phase $\varphi \triangleq -2\pi f_{RF}\tau$. The phase $\varphi$ is assumed to be constant within one integration interval. The complex Gaussian noise, $n_k$, is modeled as a sampled circular complex Gaussian white noise with variance $2\sigma^2$ as

$$\langle n_k \bar{n}_m \rangle = 2\sigma^2 \delta_{k,m}, \langle n_k n_m \rangle = 0, \langle \bar{n}_k \bar{n}_m \rangle = 0 \tag{5}$$

where $\delta_{k,m}$ is the Dirac delta function and $\langle \cdot \rangle$ represents expected value of the bracketed item with respect to the noise pdf. The signal and noise are bandlimited to $\pm f_s/2$ Hz. Therefore, the signal-to-noise ratio (SNR) is defined as

$$SNR = \frac{A^2}{2\sigma^2} \tag{6}$$

which means its carrier-to-noise density ratio ($C/N_0$) is

$$C/N_0 = SNR/f_s = \frac{A^2}{2\sigma^2 f_s} \tag{7}$$

herein. The amplitude $A$, first appearing in (4) mainly summarizes the code correlation result, baseband power $C$, and impacts of presence of data symbol; we will examine this signal-level term shortly and build a sketch view of the order of magnitude it takes by using SNR and $\sigma^2$. This completes our definition of the signal and noise models. The remaining question is how weak the target signal is that we are estimating? The answer is directly related to the necessity of developing a bound and in particular, a tighter bound in the transition and even extremely low SNR region.

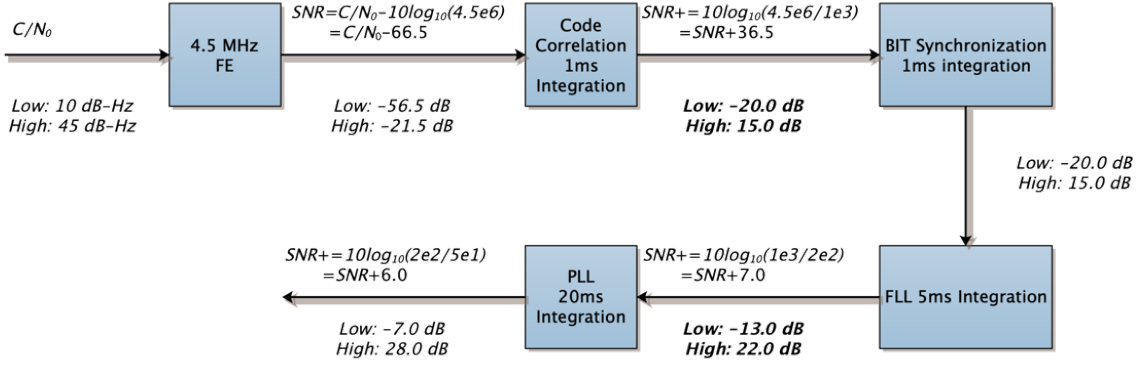

**FIGURE 2** A typical GNSS receiver processing gain example.

SNR is related to input C/N0 by input bandwidth and coherent integration time. With the assumed working C/N0 ranges from 10 to 45 dB-Hz and for a typical baseband processing chain with 4.5 MHz intermediate frequency (IF) bandwidth (Joseph, 2010, Figure 2) plotted in FIGURE 2, this means the SNR varies between -13 ~ 22 dB at the input of phase-locked loop (PLL) if the coherent integration time is 5 ms and -20 ~ 15 dB at the input of frequency locked loop (FLL) if the coherent integration time is 1 ms. Therefore, for our current problem of frequency estimation, we assume that SNR varies between -20 and 15 dB. Within this range, the transition does happen and this justifies the necessity to seek an estimation lower bound tighter than current bounds shown in FIGURE 1(b) for a wide range of GNSS applications subject to weak signals.

## 2.3 Problem Formulation

To simplify symbolization, we denote *normalized digital frequency* as

$$\theta \triangleq 2\pi f_{IF}/f_{INT} \tag{8}$$

which varies between $[-\pi, \pi]$ radians and finally, we frame our estimation problem as developing a Weiss Weinstein bound (WWB) on estimating $\theta$ using the samples

$$x_k = Ae^{i(\theta \cdot k + \varphi)} + n_k \quad k = 0, 1, \ldots, K-1 \tag{9}$$

where $A$ is determined using preset $SNR$ and $\sigma^2$ in (6); for purpose of theoretical investigation, $\sigma^2$ is set unity. A few more words about what normalization of $f_{IF}$ by $f_{INT}$ involved in $\theta$ means. Take GPS L1 C/A as an example, the primary code period is 1ms and thus $f_{INT}$ is 1 kHz. This means that the extreme $f_{IF}$ that can be estimated, usually called pull-in range, is $\pm 500$ Hz, corresponding to $\theta = \pm \pi$. Of course, for low SNR, $K > 1$ integrations should be summed up coherently to achieve a reasonable accuracy of frequency estimation. For $1/f_{int}$=5 ms, the estimable $f_{IF}$, becomes $\pm 100$ Hz (Kaplan, 2006, pp. 170), also corresponding to $\theta = \pm \pi$. Hence, normalization facilitates different integration period used in GNSS. Note that we aim to develop an error bound for MAP estimation without having to go through the intimidating calculation of the vanilla MAP estimator; 'vanilla' means no approximations are applied and therefore, we are developing a bound for the full allowable range (Kaplan, 2006, pp.170), i.e. $[-f_{INT}/2, f_{INT}/2]$; approximated such as differential schemes of MAP or ML usually means the pull-in range halved or even quartered.

## 2.4 A Priori Distribution of Circular Frequency: von Mises Distribution

The *a priori* pdf of the circular frequency is assumed to be von Mises. This distribution was studied in directional statistics and is a close approximation to wrapped normal distribution, which is in turn the circular/periodic analog of the normal distribution (Mardia, 1972). Therefore, it is particularly appropriate to characterize the parameter of interest - circular frequency since it is wrapped around $[-\pi, \pi]$ (Nitzan, 2016). The pdf of von Mises distributed parameter, $\theta$, is given as (Fisher, 1993)

$$p(\theta) = \begin{cases} \dfrac{e^{\kappa \cos(\theta - \mu)}}{2\pi I_0(\kappa)}, & \theta \in [-\pi, \pi] \\ 0, & elsewhere. \end{cases} \quad (10)$$

where $I_0(\kappa)$ is the modified Bessel function of the first kind of order zero (Mardia, 1999, pp.36); $\mu$ and $\kappa$ are the mean and concentration parameters, respectively. In particular, large $\kappa$ means smaller dispersion; when $\kappa = 0$ this distribution reduces to a uniform distribution bounded between $[-\pi, \pi]$. FIGURE 3(a) illustrates von Mises pdf subject to $\mu = 0$ and a bunch of $\kappa$ values (i.e. 0, 1, 2, 5, 20) meaningful for a typical GPS L1 C/A use case, i.e. $f_{INT} = 1$ ms and $f_{IF} \in [-500, 500]$ Hz. FIGURE 3(b) makes one step further to show the mixed effects of mean parameter $\mu$ and concentration parameter $\kappa$.

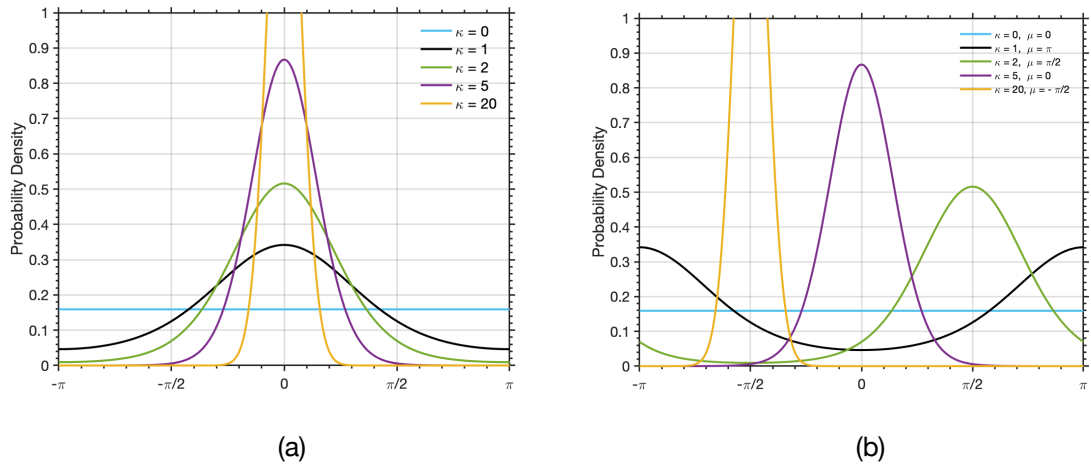

**FIGURE 3** Probability distribution function (pdf) of von Mises subject to the two defining parameters, i.e. mean $\mu$ and concentration parameter $\kappa$, for a typical GPS L1 C/A use case, i.e. $f_{INT} = 1$ ms and $f_{IF} \in [-500, 500]$ Hz. (a) $\mu = 0$ and $\kappa \in \{0, 1, 2, 5, 20\}$. (b) $(\mu, \kappa) \in \{(0,0), (\pi, 1), (\pi/2, 2), (0, 5), (-\pi/2, 20)\}$.

Note that von Mises distribution $p(\theta)$ becomes a close approximation to a wrapped normal distribution with parameters $\mu$ and $1 - I_1(\kappa)/I_0(\kappa)$ as its defining parameters when $\kappa$ approaches infinity (Mardia, 1999, pp.38), where $I_1(\kappa)$ is the modified Bessel function of the first kind of order one. However, other approximations exist. For the current problem where $\kappa$ seldom exceeds 20, von Mises can be approximated by a normal distribution with mean $\mu$ and a variance equal to $-2\ln(I_1(\kappa)/I_0(\kappa))$ (Berens, 2009).

## 3. WEISS-WEINSTEIN BOUND FOR GNSS FREQUENCY ESTIMATION

Signals with changing parameters are often encountered in many practical signal processing applications and endorse many ways of characterization. Such changes, known in literature in a broader sense as 'change-points' (Bacharach, 2017), are often considered random and can be reflected by shifts in the parameters of its distribution. Among all the possible estimation schemes, ML or MAP estimators often preferred for their good statistical properties, easy circuit implementation and most important of all, optimum performance in appropriate conditions. However, in the context of change-point estimation, certain regularity assumptions, usually used to prove the asymptotic normality of the ML or MAP estimators, are obviously not fulfilled, this is especially due to the discrete nature of the parameters of interest. Consequently, the study of such estimator performance requires a specific analysis. One of the accessible ways of closed-form performance evaluation is to use lower bounds with less stringent requirements on regularity and as we just mentioned in the introduction, WWB is an exact example of this family with less regularity conditions and in particular a lower bound which does not require the differentiability of its likelihood.

In this section, Weiss Weinstein bound (WWB) is developed and its link to other modern bounds is explained.

For an arbitrary estimation problem with an observation vector $\mathbf{x}$ and parameter vector $\boldsymbol{\theta}$ of length $D$, the generic WWB is (Weiss & Weinstein, 1985; Weinstein & Weiss, 1988)

$$\max_{s_1,\ldots,s_R,\mathbf{h}_1,\ldots,\mathbf{h}_R} \mathbf{H}\mathbf{Q}^{-1}\mathbf{H}^T \tag{11}$$

where $\mathbf{Q} \in \mathbb{R}^{R \times R}$ whose element will be defined shortly and $\mathbf{H}$ is defined as

$$\mathbf{H} = [\mathbf{h}_1, \mathbf{h}_2, \ldots, \mathbf{h}_R] \tag{12}$$

where each $\mathbf{h}_i \in \mathbb{R}^{D \times 1}$ and therefore $\mathbf{H} \in \mathbb{R}^{D \times R}$. $R$ is a chosen parameter indicating the total number of test points $\mathbf{h}_i$ whose selection method will be dealt with in Section 4.3. Intuitively, the appealing aspect of WWB comes from the fact that CRB is evaluated only around the (assumed) best estimate and thus has only one test point, while WWB is evaluated over a much wider uncertainty range of the parameter space which can be far from the best estimate, thus grasping large errors while SNR is small. An element of matrix $\mathbf{Q}$ with $i$ and $j$ as its row and column index, respectively, is defined as

$$[\mathbf{Q}]_{ij} = \frac{E_{\mathbf{x},\boldsymbol{\theta}}\{[L^{s_i}(\mathbf{x};\boldsymbol{\theta}+\mathbf{h}_i,\boldsymbol{\theta}) - L^{1-s_i}(\mathbf{x};\boldsymbol{\theta}-\mathbf{h}_i,\boldsymbol{\theta})][L^{s_j}(\mathbf{x};\boldsymbol{\theta}+\mathbf{h}_j,\boldsymbol{\theta}) - L^{1-s_j}(\mathbf{x};\boldsymbol{\theta}-\mathbf{h}_j,\boldsymbol{\theta})]\}}{E_{\mathbf{x},\boldsymbol{\theta}}\{L^{s_i}(\mathbf{x};\boldsymbol{\theta}+\mathbf{h}_i,\boldsymbol{\theta})\}E_{\mathbf{x},\boldsymbol{\theta}}\{L^{s_j}(\mathbf{x};\boldsymbol{\theta}+\mathbf{h}_j,\boldsymbol{\theta})\}} \tag{13}$$

$$0 \leq i,j \leq R$$

where $0 < s_i < 1$, $0 < s_j < 1$, and function $L(\mathbf{x};\boldsymbol{\theta}_1,\boldsymbol{\theta}_2)$ is the joint likelihood ratio with

$$L(\mathbf{x};\boldsymbol{\theta}_1,\boldsymbol{\theta}_2) = \frac{p(\mathbf{x},\boldsymbol{\theta}_1)}{p(\mathbf{x},\boldsymbol{\theta}_2)} \tag{14}$$

as its definition. For our problem, only one parameter is to be estimated and thus the parameter vector $\boldsymbol{\theta}$ reduces to a scalar $\theta$ and each column of $\mathbf{h}$ now reduces to a scalar $h_i$, and $\mathbf{H} = [h_1, h_2, \ldots, h_R]$ is now a row vector of length $R$. With these in mind, after expanding the nominator of $\mathbf{Q}$, we arrive at

$$[\mathbf{Q}]_{ij}$$

$$= \frac{E_{\mathbf{x},\theta}\{[L^{s_i}(\mathbf{x};\theta+h_i,\theta)-L^{1-s_i}(\mathbf{x};\theta-h_i,\theta)][L^{s_j}(\mathbf{x};\theta+h_j,\theta)-L^{1-s_j}(\mathbf{x};\theta-h_j,\theta)]\}}{E_{\mathbf{x},\theta}\{L^{s_i}(\mathbf{x};\theta+h_i,\theta)\}E_{\mathbf{x},\theta}\{L^{s_j}(\mathbf{x};\theta+h_j,\theta)\}}$$

$$= \frac{\left[E_{\mathbf{x},\theta}\left\{\frac{p^{s_i}(\mathbf{x},\theta+h_i)p^{s_j}(\mathbf{x},\theta+h_j)}{p^{s_i+s_j}(\mathbf{x},\theta)}\right\} - E_{\mathbf{x},\theta}\left\{\frac{p^{s_i-s_j-1}(\mathbf{x},\theta)p^{s_j}(\mathbf{x},\theta+h_j)}{p^{s_i-1}(\mathbf{x},\theta-h_i)}\right\} \right.}{E_{\mathbf{x},\theta}\left\{\frac{p^{s_i}(\mathbf{x},\theta+h_i)}{p^{s_i}(\mathbf{x},\theta)}\right\} \cdot E_{\mathbf{x},\theta}\left\{\frac{p^{s_j}(\mathbf{x},\theta+h_j)}{p^{s_j}(\mathbf{x},\theta)}\right\}} \quad (15)$$

$$\left. \frac{-E_{\mathbf{x},\theta}\left\{\frac{p^{s_j-s_i-1}(\mathbf{x},\theta)p^{s_i}(\mathbf{x},\theta+h_i)}{p^{s_j-1}(\mathbf{x},\theta-h_j)}\right\} + E_{\mathbf{x},\theta}\left\{\frac{p^{s_i+s_j-2}(\mathbf{x},\theta)}{p^{s_i-1}(\mathbf{x},\theta-h_i)p^{s_j-1}(\mathbf{x},\theta-h_j)}\right\}\right]}{}$$

and the **Q** matrix can be evaluated using *a priori* and *a posteriori* pdf 's.

### 3.1 Weiss-Weinstein Bound for GNSS Frequency Estimation

In Appendix A, Weiss Weinstein bound is developed in detail for frequency estimation in a generic GNSS setting. In short, $[\mathbf{Q}]_{ij}$ can be computed as

$$[\mathbf{Q}]_{ij} = \frac{e^{\mu_{ij,1}+\gamma_{ij,1}} - e^{\mu_{ij,2}+\gamma_{ij,2}} - e^{\mu_{ij,3}+\gamma_{ij,3}} + e^{\mu_{ij,4}+\gamma_{ij,4}})}{e^{\mu_i+\gamma_i}e^{\mu_j+\gamma_j}} \quad (16)$$

where exponents $\mu_{ij,n}$, $\gamma_{ij,n}$, $n \in \{1,2,3,4\}$, and $\mu_i$ and $\gamma_i$ are defined by

$$\mu_{ij,1} = SNR\left[K\left((s_i+s_j-1)^2+s_i^2+s_j^2-1\right)+2s_is_j\frac{\cos\left[\frac{(h_i-h_j)(K-1)}{2}\right]\sin\left[\frac{(h_i-h_j)K}{2}\right]}{\sin\left[\frac{(h_i-h_j)}{2}\right]}\right.$$

$$\left. -2(s_i+s_j-1)s_i\frac{\cos\left[\frac{h_i(K-1)}{2}\right]\sin\left[\frac{h_iK}{2}\right]}{\sin\left[\frac{h_i}{2}\right]} - 2(s_i+s_j-1)s_j\frac{\cos\left[\frac{h_j(K-1)}{2}\right]\sin\left[\frac{h_jK}{2}\right]}{\sin\left[\frac{h_j}{2}\right]}\right] \quad (17)$$

$$\mu_{ij,2} = SNR\left[K\left(s_j^2+(s_i-1)^2+(s_i-s_j)^2-1\right)-2s_j(s_i-1)\frac{\cos\left[\frac{(h_i+h_j)(K-1)}{2}\right]\sin\left[\frac{(h_i+h_j)K}{2}\right]}{\sin\left[\frac{(h_i+h_j)}{2}\right]}\right.$$

$$\left. +2s_j(s_i-s_j)\frac{\cos\left[\frac{h_j(K-1)}{2}\right]\sin\left[\frac{h_jK}{2}\right]}{\sin\left[\frac{h_j}{2}\right]} - 2(s_i-1)(s_i-s_j)\frac{\cos\left[\frac{h_i(K-1)}{2}\right]\sin\left[\frac{h_iK}{2}\right]}{\sin\left[\frac{h_i}{2}\right]}\right] \quad (18)$$

$$\mu_{ij,3} = SNR\left[K\left(s_i^2+(s_j-1)^2+(s_i-s_j)^2-1\right)-2s_i(s_j-1)\frac{\cos\left[\frac{(h_i+h_j)(K-1)}{2}\right]\sin\left[\frac{(h_i+h_j)K}{2}\right]}{\sin\left[\frac{(h_i+h_j)}{2}\right]}\right.$$

$$\left. +2s_i(s_j-s_i)\frac{\cos\left[\frac{h_i(K-1)}{2}\right]\sin\left[\frac{h_iK}{2}\right]}{\sin\left[\frac{h_i}{2}\right]} - 2(s_j-1)(s_j-s_i)\frac{\cos\left[\frac{h_j(K-1)}{2}\right]\sin\left[\frac{h_jK}{2}\right]}{\sin\left[\frac{h_j}{2}\right]}\right] \quad (19)$$

$$\mu_{ij,4} = \text{SNR}\left[K\left((s_i+s_j-1)^2 + (s_i-1)^2 + (s_j-1)^2 - 1\right) - 2(s_i+s_j-1)(s_i-1)\frac{\cos\left[\frac{h_i(K-1)}{2}\right]\sin\left[\frac{h_iK}{2}\right]}{\sin\left[\frac{h_i}{2}\right]}\right.$$
$$\left.-2(s_i+s_j-1)(s_j-1)\frac{\cos\left[\frac{h_j(K-1)}{2}\right]\sin\left[\frac{h_jK}{2}\right]}{\sin\left[\frac{h_j}{2}\right]} + 2(s_i-1)(s_j-1)\frac{\cos\left[\frac{(h_i-h_j)(K-1)}{2}\right]\sin\left[\frac{(h_i-h_j)K}{2}\right]}{\sin\left[\frac{(h_i-h_j)}{2}\right]}\right] \quad (20)$$

and

$$\gamma_{ij,1} = \ln\int_{-\pi}^{\pi-h_i}\frac{1}{2\pi I_0(\kappa)}e^{\kappa[(1-s_i-s_j)\cdot\cos(\theta-\mu)+s_i\cdot\cos(\theta+h_i-\mu)+s_j\cdot\cos(\theta+h_j-\mu)]}d\theta \quad (21)$$

$$\gamma_{ij,2} = \ln\int_{-\pi+h_j}^{\pi-h_i}\frac{1}{2\pi I_0(\kappa)}e^{\kappa[(s_i-s_j)\cdot\cos(\theta-\mu)+s_j\cdot\cos(\theta+h_j-\mu)+(1-s_i)\cdot\cos(\theta-h_i-\mu)]}d\theta \quad (22)$$

$$\gamma_{ij,3} = \ln\int_{-\pi+h_i}^{\pi-h_j}\frac{1}{2\pi I_0(\kappa)}e^{\kappa[(s_j-s_i)\cdot\cos(\theta-\mu)+s_i\cdot\cos(\theta+h_i-\mu)+(1-s_j)\cdot\cos(\theta-h_j-\mu)]}d\theta \quad (23)$$

$$\gamma_{ij,4} = \ln\int_{-\pi+h_i}^{\pi}\frac{1}{2\pi I_0(\kappa)}e^{\kappa[(s_i+s_j-1)\cdot\cos(\theta-\mu)+(1-s_i)\cdot\cos(\theta-h_i-\mu)+(1-s_j)\cdot\cos(\theta-h_j-\mu)]}d\theta \quad (24)$$

and

$$\mu_i = -s_i(1-s_i)\cdot 2K\cdot SNR\left(1 - \frac{1}{K}\cos\left[\frac{h_i(K-1)}{2}\right]\cdot\frac{\sin\left[\frac{h_iK}{2}\right]}{\sin\left[\frac{h_i}{2}\right]}\right) \quad (25)$$

and

$$\gamma_i = \ln\int_{-\pi}^{\pi-h_i}\frac{1}{2\pi I_0(\kappa)}e^{\kappa[(1-s_i)\cdot\cos(\theta-\mu)-s_i\cdot\cos(\theta+h_i-\mu)]}d\theta \quad (26)$$

respectively. $\mu_j$ and $\gamma_j$ is obtained by just replacing subscript $i$ with $j$ in $\mu_i$ and $\gamma_i$ accordingly.

### 3.2 Links to Other Bounds

Some other bounds, for example, Bayesian Cramér-Rao bound (BCRB) and Bobrovsky-Zakai bound (BZB), can be obtained as a special case of Weiss Weinstein bound. BCRB is obtained with $R = 1$ and $s_i \to 1^-$ and $s_j \to 1^-$ when $h_i \to 0$ and $h_j \to 0$ (Weinstein & Weiss, 1988). BZB is obtained when $s = 1$. Specifically, the Bayesian information matrix (BIM) $\mathbf{J}_B$ for an observation vector $\mathbf{x}$ of length $K$ and parameter vector $\boldsymbol{\theta}$ of length $D$ is (Van Trees, 1968)

$$[\mathbf{J}_B]_{ij} = E_{\mathbf{x},\boldsymbol{\theta}}\left\{\frac{\partial \ln p(\mathbf{x},\boldsymbol{\theta})}{\partial \theta_i}\cdot\frac{\partial \ln p(\mathbf{x},\boldsymbol{\theta})}{\partial \theta_j}\right\} \quad (27)$$

and the BRCB for this problem is thus defined as

$$\mathbf{J}_B^{-1} = BCRB \quad (28)$$

readily. After some substitutions and manipulations, BCRB for this problem can be formulated as

$$\mathbf{J}_B = SNR \frac{K(K-1)(2K-1)}{3} + \kappa \cdot \frac{I_1(\kappa)}{I_0(\kappa)} \quad (29)$$

and its derivation is provided in Appendix B. We will use BCRB as one of the two benchmarks (the other is Ziv-Zakai bound) in the simulations to highlight the bounding performance of the developed bound.

## 4. TWO-STEP OPTIMIZATION OF WEISS WEINSTEIN BOUND

The advantage of WWB lies in its power to predict the threshold region of RMSE while CRB or Bayesian CRB cannot. The developed bound is a maximization of the matrix $\mathbf{HQ}^{-1}\mathbf{H}^T$ over a combination of two sets of parameters, $\{h_i\}$ and $\{s_i\}$, $i = 1, ..., R$, as in Equation (10). The row vector $\mathbf{H} = (h_1, ..., h_R)$ is often called 'test points' in the literature (Weinstein, 1988). Test points are estimated parameters supposed to be outliers that lead to large RMSE. Therefore, evaluation of WWB is also an optimization procedure. For estimating a normalized frequency in GNSS, test points take values between $[-\pi, \pi]$. A typical value of the vector size, $R$, can easily exceed 10 (single sided, Weinstein, 1988), which means we are facing an optimization problem including about 20 parameters (two sided) to be optimized over, a quite intimidating task. Fortunately, we have some rules-of-thumb (Weinstein, 1988; Brown & Liu, 1993; Bell et al., 1997) to follow. We propose that $\{h_i\}$ and $\{s_i\}$ need not be optimized simultaneously. In the sequel, we will solve this maximization problem using a divide and conquer strategy. Firstly, we fix $\mathbf{H}$ and optimize over $\{s_i\}$; then we use the optimized $\{s_i\}$ to further optimize the bound over $\mathbf{H}$. But, before all these happen, it is helpful to explain the reasons why the WWB is better than other Bayesian bounds such as BCRB in predicting the threshold. The explanation helps in understanding the fundamental problem of optimization, i.e., how to select test points.

### 4.1 Test Point: The Reason Why WWB Outperforms BCRB

As mentioned in section 3.2, BCRB can be obtained from WWB if the length of the test point vector is 1 and the single test point, $h_1$, approaches to zero. Therefore, we will use BCRB as an illustrative example for its simplicity. Suppose now we have a noisy signal and let us plot the logarithm of its *a posteriori* probability

$$A_x(\theta) = (2K)(SNR)\Re e \left\{ e^{-j\varphi} \frac{1}{K} \sum_{k=0}^{K-1} x_k \, e^{-j\theta \cdot k} \right\} + \kappa \cdot \cos(\theta - \mu) \quad (30)$$

in FIGURE 4 to illustrate the behavior of MAP estimates, which the bounds we discussed so far are trying to bound. Parts of the log-likelihood that does not depend on $\hat{\theta}$ is deliberately omitted and here $\kappa = 0$ is assumed without loss of generality. These simplifications are solely for better visual effects, since logarithm of *a posteriori* probability differs from the log-likelihood by a varying, i.e. not constant "noise floor" dictated by the *a priori* distribution. This noise floor will blur the main lobes and side lobes of the ambiguity surface once the noise gets higher, thus hiding BCRB's characteristic dependence on side lobe positions. As a side note, if the *a priori* distribution is uniform, the *a posteriori* log-likelihood reduces to log-likelihood because the noise floor is simply lifted by a constant. We will stick to the log-likelihood for the current discussion.

The reason why WWB outperforms BCRB lies in the choice of test vector: the number and locations of points. BCRB use a single test point extremely close to the main lobe while WWB use multiple test points, including the one used for BCRB computation. FIGURE 4(a) shows a noise-free ambiguity surface and in (b) we show three realizations of the corresponding noisy ambiguity surface for three values of SNR and $K = 20$ which means the integration time is now 20 ms for GPS L1 C/A. For $\text{SNR} = 5 \text{ dB}$, the main lobe peaks are always correctly picked by the maximization and indexing operation, but for increasing levels of noises, the same operation tends to pick wrong peaks. This is the reason underlying the RMSE's thresholding behavior. During the transition of SNR from medium to low values, BCRB's choice of a single test point is slow to reflect this change in the log-likelihood function, since sometimes, the peak is still correctly picked, as shown in the second plot of FIGURE 4(b), when 2 out of 3 trials succeed in estimating the true value.

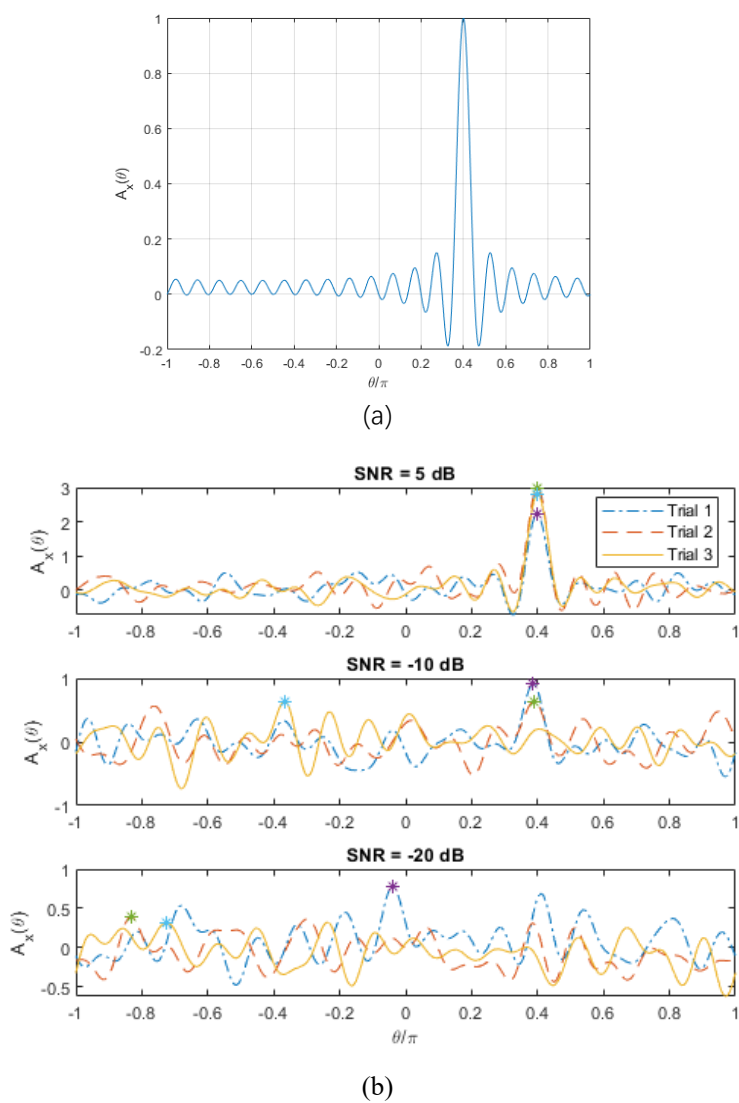

**FIGURE 4** Ambiguity surface $A_x(\theta)$ with $K = 20$. True frequency is set as $0.4\pi$. Known phase $\varphi = \pi/6$ (a) $A_x(\theta)$ with $x_k = r_k$. (b) Noisy $A_x(\theta)$ with $x_k = r_k + n_k$

In contrast, WWB uses multiple test points such that it has fewer chances of missing the gross errors or incorrect estimates, but still, the locations and number of test points need optimizing and the next two subsections are dedicated to optimization over WWB's two hyper parameters, namely the exponential index $\{s_i\}$ and the test points $\{h_i\}$ where $i = 1, \ldots, R$; $R$ is an arbitrary positive integer. Historically, selection of test points appears in prior art as assertions and several rules of thumb are obtained for specific applications. We will extend this trial-and-error way, but deal with the exponential index and test point positions separately to find an optimum set of test points in section 4.3. We will use the legacy test points in section 4.2 to optimize WWB over $\{s_i\}$, as the first step of the two-step optimization procedure.

**4.2 Step 1: Optimization over $\{s_i\}$**

Previous work about WWB left the choice of $\{s_i\}$ almost untouched and only an arbitrary value such as 0.5 is magically provided (Weinstein & Weiss, 1988), although for such a simple problem with only a single test point, i.e. $R = 1$, it is easy to prove that $s = 0.5$ is indeed the maximizer. Aiming for a more general case, we propose to use a brute-force method by numerically evaluating WWB to fix a vector of $s_i$ that maximizes WWB. As a first step, we will use fixed, i.e. legacy test points (Xu, 2001; DeLong, 1993; Xu et al., 2004). These test points are composed of:

(1) small values extremely close to the main lobe peak, e.g. $h_i = 0.01\pi$, $0.001\pi$, such that $\theta + h_i$ will be very close to the main lobe peak position, $\theta$; this is to ensure WWB behaves correctly at high SNR (or asymptotic region), i.e. WWB will converge just like BCRB and also, CRB in this region;

(2) frequency points corresponding to the side lobe peaks of the $A_x(\theta)$; for our problem we use all frequency points corresponding to positive side lobe peaks between $[0, \pi]$. This is to ensure WWB's behavior in medium-to-low, i.e. thresholding and low SNR regions.

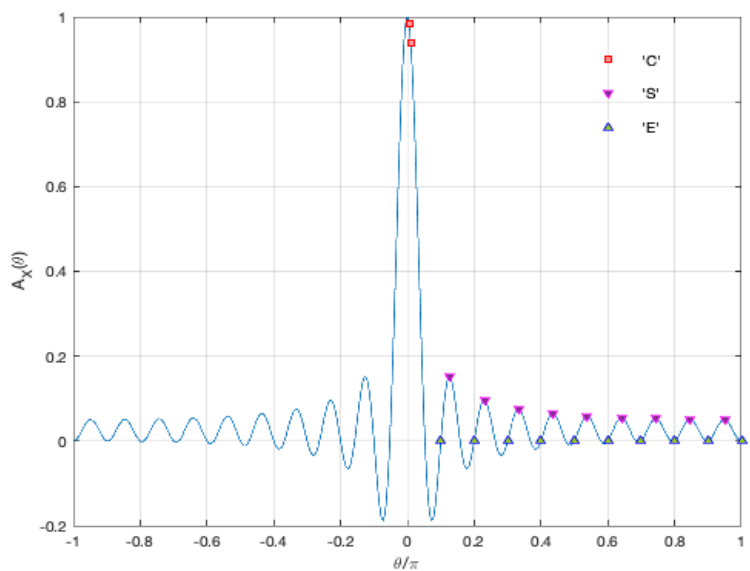

**FIGURE 5** Legacy and proposed test points. Legacy test points compose only 'C' and 'S' points; the proposed test points include 'E' points while retaining all 'C' and 'S' points.

Again, we use a noise-free MAP estimation function $A_x(\theta)$ from Equation (30) as an illustrative example. The legacy test points (groups 'C' and 'S') are shown in FIGURE 5.

With these test points, solution of WWB, formulated as Equation (11), is now an optimization problem only over $\{s_i\}$, i.e.

$$\max_{s_1,\ldots,s_R} \mathbf{H}\mathbf{Q}^{-1}\mathbf{H}^T \tag{31}$$

and as mentioned before, this could lead to a tough problem because the total number of parameters in the end could be as high as $R^2$, where $R$ is the total number of test points (type 'C', 'S', and 'E'), eventually rendering the optimization intractable. Therefore, we adopt a simplified assumption:

$$s_i = s, i = 1, \ldots, R \tag{32}$$

i.e., $s_i$'s are all equal-valued. Now the only problem is to select a value within range of $(0,1)$. Since derivatives of WWB is almost analytically impossible, we use numerical ways by comparing WWB with different $s$ values. Specifically, $s \in \{0.1,0.2,0.3,0.4,0.5,0.6,0.7,0.8,0.9\}$ is tried. Results show that (1) $s = 0.5$ is the maximizer and (2) $s$ and (1-s) produce identical WWB's. An example of this result is shown in FIGURE 6. For better visualization, only WWB's with $s = 0.1$ and $s = 0.5$ are shown.

Although here the *a priori* pdf is parameterized with a particular choice, $\mu = 0$ and $\kappa = 2$, the result is also true for other combinations of $\mu$ and $\kappa$. We defer the discussion of how *a priori* pdf affects the developed WWB to section 5. In the following sections, $s_i = s = 0.5$ will be assumed.

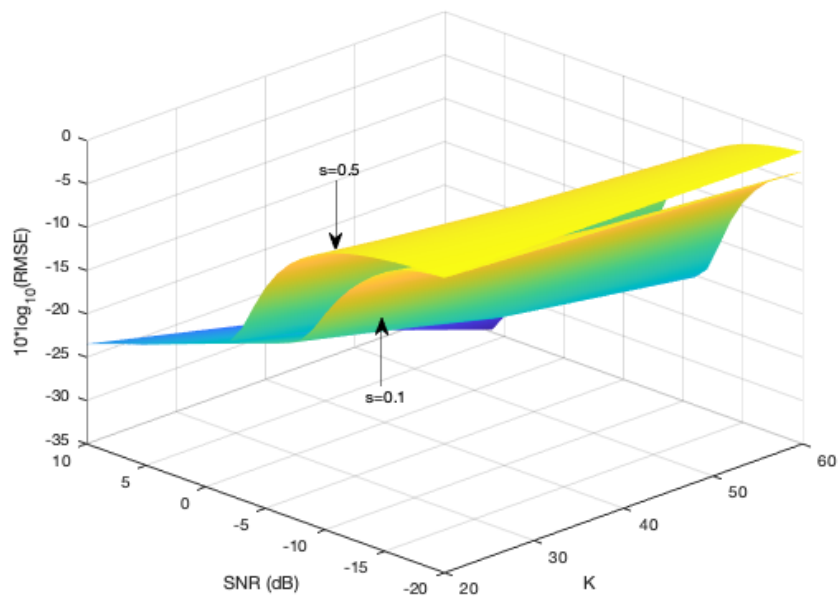

**FIGURE 6** WWB with $s = 0.1$ and $s = 0.5$. Sample numbers, $K \in \{20,40,60\}$ and $SNR \in [-20,10]$ (dB). Von Mises distribution: $\mu = 0$, $\kappa = 2$.

## 4.3 Step 2: Optimization over Test Point Vector, h

As the second step in the two-step optimization procedure, this subsection deals with optimization over test points $\{h_i\}$ and presents the proposed test points.

The proposed test points are shown in FIGURE 5. We propose a trio of arguments, $(C, S, E)$, to describe the configuration of WWB's test point: 'C' is the number of points <u>c</u>lose to the main lobe peak, 'S' the number of points corresponding to <u>s</u>ide lobe peaks, and 'E' the number of points <u>e</u>venly spaced between $[0.1\pi, \pi]$. The 'E' component is the additional improvement we propose to append to the legacy test point vector. Following the thinking that more points lead to better performance (Xu et al., 2004; DeLong, 1993; Xu, 2001), the proposed test points include:

(1) small values extremely close to the main lobe peak; this part is identical to the legacy ones,

(2) bunch of side lobe peaks selected exactly as those included in the legacy vector, and

(3) an extra vector evenly spaced between $[0.1\pi, \pi]$.

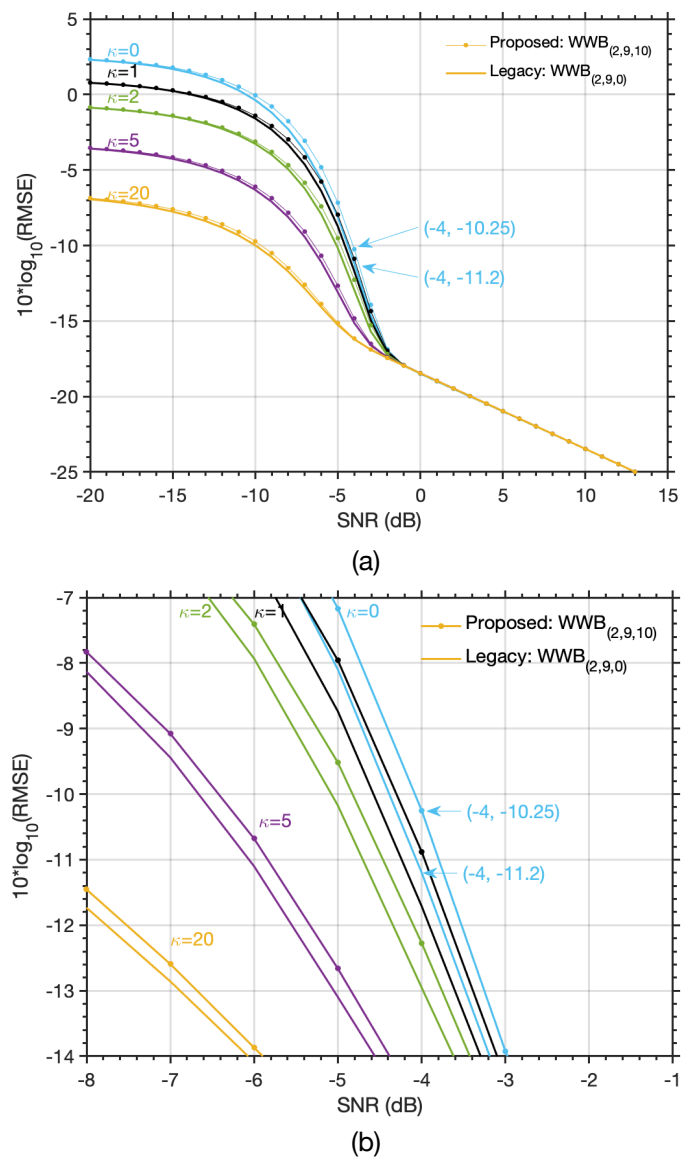

**FIGURE 7** (a) WWB: Legacy V.S. proposed configurations. $K = 20$. $\mu = 0$. The maximal improvement occurs at SNR = -4 dB and is 0.95 dB. (b) Zoom-in of (a) around $\text{SNR} = -4\text{dB}$.

Points (1) and (2) are exactly the legacy points in section 4.2. Reusing (1) from the legacy points is due to the fact that these two points are enough for a good asymptotic behavior, and reusing (2) from the legacy points is because frequency points corresponding to side lobe peaks bring the biggest performance boost with least number of additional points.

As an example, FIGURE 7(a) shows an example of WWB using the legacy and proposed test points when $K = 20$ and increasing $\kappa$ values of von Mises pdf. With the introduced trio notation, in FIGURE 7(a), $\text{WWB}_{(2,9,0)}$ is the legacy configuration and $\text{WWB}_{(2,9,10)}$ is the proposed configuration. The evenly spaced points proposed is aimed to provide additional chances of capturing 'wild' peaks during mid-to-low SNR transition, thus increasing the value of WWB in this transition region. The improvement is 0.95 dB at $\text{SNR} = -4$ dB, equivalent to a 19.8 % $\left(= 10^{\frac{0.95}{10}}/(2\pi)\right)$ increase in predicted RMSE. Take GPS L1 C/A for example, this means a 3 Hz improvement in bounding accuracy: $f_{INT} \cdot |RMSE_1 - RMSE_2|/(2\pi)$, where $f_{INT} = 1000$ Hz. Note here $RMSE_i (i = 1,2)$ represents the RMSE's of the two compared bounds and should be converted from dB to radians first. This performance improvement can be better perceived with a close-up, i.e. FIGURE 7(b). The reason to stop continuing adding in evenly spaced points is that with increasing number of points, the performance margin gets smaller and smaller and finally diminishes.

## 5. SIMULATIONS

This section illustrates the bounding performance of WWB with the proposed model parameters through numerical simulation. The simulations consider a typical GNSS receiver architecture presented in FIGURE 2 indicating SNR varies between -20 to 15 dB for frequency estimation.

### 5.1 Numerical Challenges

Computation of WWB, as evident from Equation (16), requires $O(R^2)$ time, where $R$ is the length of the test point vector and $O$ represents the big O notation of computational complexity in time (in fact, also in space). This prevents a quick evaluation of WWB. For large inference network using signal bound as input (Manlaney, 2020), evaluating WWB in a quick and simple manner is still what we are after. Therefore, a reasonable number of test points are needed and it is meaningful to investigate whether all the positive side lobe peak points should be included in the test point vector. This is especially important when the data length, $K$, which roughly two times the number of side lobe peaks, increases to a large number of for example, 1000 for extremely low SNR conditions. $K$ was assumed to be 20 in the previous examples and will still be used here to help select the appropriate number of side lobe peaks. We are not forced to use all these points, so the nine positive side lobe points, starting from the point closest to the main lobe peak, are progressively added to the test vector and the results are shown in FIGURE 8(a)-(f), with different combinations of $\mu$ and $\kappa$.

To examine only the effects of the side lobe peaks and for better visualization, FIGURE 8 compares $\text{WWB}_{(2,n,0)}$ where $n \in \{1,3,5,7,9\}$, i.e. only half of the test points, to the legacy configuration, $\text{WWB}_{(2,9,0)}$. In each case, increasing number of test points help to obtain a tighter bound, but this benefit becomes less significant with the increase of $\kappa$ (FIGURE 8(a)-(c)). We conclude that a properly small value of the concentration parameter of von Mises distribution, i.e., assuming frequency distributes like a flatter bell shape suffices. Changes of

$\mu$ seldom leads to changes of bound values (FIGURE 8(d) v.s. (e), and also, FIGURE 8(b) v.s. (f)). With all these in mind, we will proceed with a parameterization of $WWB_{(2,9,10)}$, i.e., 2 points close to the main lobe peak, 9 side lobe peak points, and 10 extra points evenly spaced between $[0.1\pi, \pi]$ in the ensuing simulations.

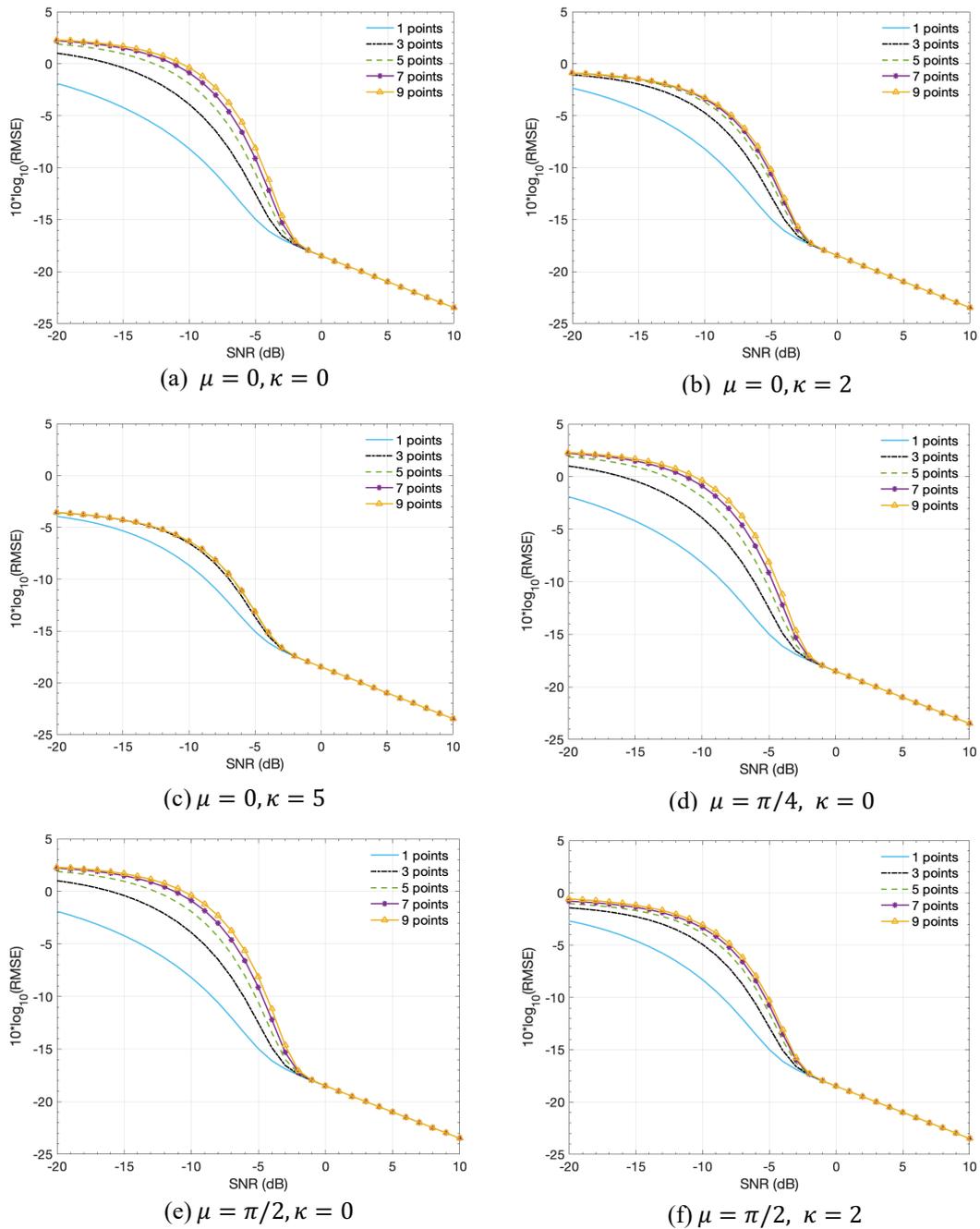

(a) $\mu = 0, \kappa = 0$

(b) $\mu = 0, \kappa = 2$

(c) $\mu = 0, \kappa = 5$

(d) $\mu = \pi/4, \ \kappa = 0$

(e) $\mu = \pi/2, \kappa = 0$

(f) $\mu = \pi/2, \ \kappa = 2$

**FIGURE 8** $\text{WWB}_{(2,n,0)}$ with increasing number of test points and different combinations of $\mu$ and $\kappa$.

## 5.2 WWB Dependence on *A Priori* Distribution

In discussing FIGURE 8, we know by observation, that the effects of two parameters of our *a priori* distribution, $\mu$ and $\kappa$ can be decoupled, but the conclusion is drawn with limited combinations of $\mu$ and $\kappa$. FIGURE 9 (a) shows the joint effects of these two parameters on the developed WWB with $\text{SNR} = 0$ dB and sample counts $K = 100$ in milliseconds. We can see that the effects of $\mu$ and $\kappa$ can be decoupled within a limited range of $\mu \in (-\frac{\pi}{2}, \frac{\pi}{2})$ and $\kappa \in (3, +\infty)$. Although theoretically, this result is true for any large values of $\kappa > 3$, $\kappa$ cannot be infinitely large since the computation of WWB will experience numerical problems for the *a priori* pdf now takes the shape of a Dirac delta function. The von Mises random samples are generated using open-source packages (Berens, 2009).

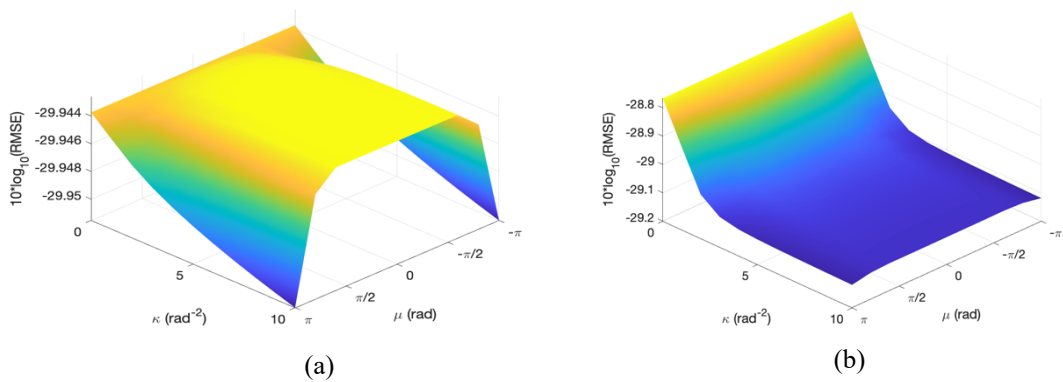

(a)            (b)

**FIGURE 9** WWB dependence on von Mises distribution with parameters $\mu$ and $\kappa$. (a) $K = 100$, $\text{SNR} = 0$ dB. (b) $K = 200$, $\text{SNR} = -10$ dB.

Though the result is obtained with specific SNR and sample count $K$, it is generally true for other SNR and $K$ values, as seen in FIGURE 9(b). Note that when $\kappa$ is small ($< 3$) and $\mu \in (-\frac{\pi}{2}, \frac{\pi}{2})$, the bounds in FIGURE 9(a) and (b) go in reverse directions. However, this divergence is negligible since the difference does not exceed 0.5 dB.

## 5.3 WWB Dependence on Input SNR and Coherent Integration Time K

Previously, FIGURE 6 shows WWB values change with input signal SNR and number of input samples, $K$. As defined in section 2.2, $K$ is the number of primary PRN code integration interval and thus is also the coherent integration length in primary PRN code length. With the observations in section 5.2 that small $\kappa$ values $(< 3)$ is of particular interest, we plot this dependence in FIGURE 10 with $\kappa = 1$ and $\mu = 0$.

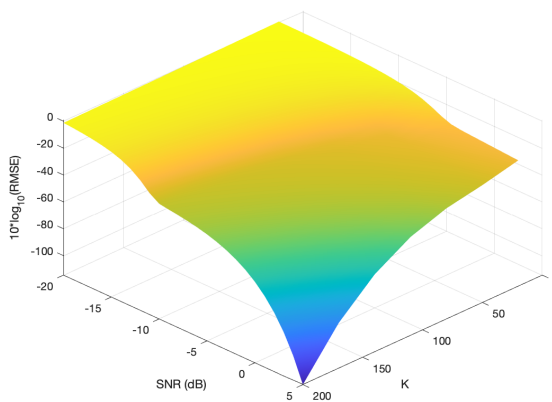

**FIGURE 10** WWB dependence on input SNR and number of primary PRN code period. $(\kappa, \mu) = (1, 0)$

The results show that larger $K$ or larger SNR always means smaller estimation errors and for weak signal deep into $-10$ dB or even lower, $K$ required to reduce estimation errors increase tremendously. This coincides with our knowledge about this type of estimators.

### 5.4 WWB Compared to Other Modern Bounds

Bayesian Cramér-Rao bound (BCRB) and Ziv-Zakaï bound (ZZB) are used as benchmarks to check whether WWB is indeed the tightest bound, in most cases, among modern bounds when GNSS frequency estimation is concerned. A brief development of BCRB is provided in Appendix B. FIGURE 11 depicts WWB against BCRB and MAP results with different $\kappa$ values while $\mu$ is defaulted to zero.

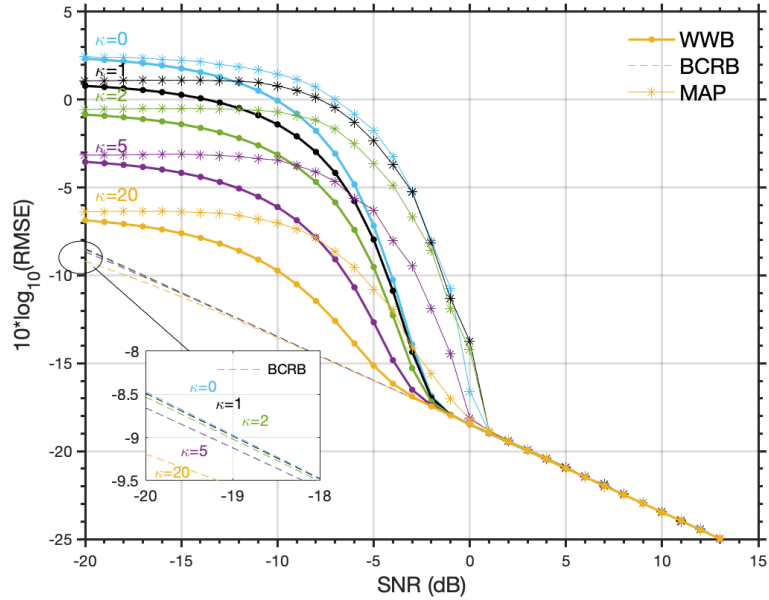

**FIGURE 11** Comparison of WWB to MAP and BCRB, with $\kappa \in \{0, 1, 2, 5, 20\}$ and $\mu = 0$.

The MAP result here is an averaged of 10,000 Monte Carlo trials. An individual run of MAP estimation can be done by

$$\hat{\theta}_{MAP} = \arg\max_{\theta}\{\ln p(\mathbf{x}, \theta)\} \tag{33}$$

where $\ln p(\mathbf{x}, \theta)$ reduces to the following form using the likelihood $p(\mathbf{x}|\theta)$ and prior distribution $p(\theta)$, i.e. von Mises as

$$\begin{aligned}\ln p(\mathbf{x}, \theta) &= \sum_{k=0}^{K-1} \ln p(x_k|\theta) + \ln p(\theta) \\ &= 2K \cdot SNR \cdot \Re e\left[e^{-j\varphi}\frac{1}{K}\sum_{k=0}^{K-1} x_k e^{-j(\theta \cdot k)}\right] + \kappa\cos(\theta - \mu) - \ln(2\pi I_0(\kappa))\end{aligned} \tag{34}$$

by assuming that the observation vector $\mathbf{x}$ consists of $K$ independent and identically distributed (i.i.d.) samples from a Gaussian distribution with zero mean and variance $\sigma^2$ as mentioned in signal model defined in Section 2 and the phase $\varphi$ is constant during the integration/summation. The square bracketed term is easily identified as the digital Fourier transform of the observation vector and the last term, $-\ln(2\pi I_0(\kappa))$, does not depend on the parameters of interest and thus can be removed in the simulation.

In can be seen in FIGURE 11, for high SNR, both BCRB and WWB converges to MAP results and for low SNR, these two bounds flatten out at the prior variance. Within the entire SNR range, WWB is a tight lower bound for MAP error. The threshold occurs for both WWB and MAP at the same SNR while BCRB fails to do so. For low SNR, we can also observe that WWB converges to MAP error. Since we are interested in SNR exceeding -20dB, we do not show comparison between WWB with BCRB or MAP in exceedingly small SNR say -20 ~ -45 dB, where WWB and BCRB eventually converges to MAP since there is no extra information provided by signal samples now; the only information the estimator can depend upon is prior knowledge of the estimated values.

As a more competent competitor, ZZB is given without proof as (Bell et al., 1997)

$$\frac{1}{2}\int_0^\infty \Lambda\{\int_{-\infty}^\infty [p(\theta) + p(\theta + h)]P_{\min}(\theta, \theta + h)d\theta\}hdh \tag{35}$$

where $P_{\min}$ denotes the minimum probability of error obtained from the optimum likelihood ratio test and $\Lambda(f(h))$ is a non-increasing function of $h$ obtained by filling in any valleys in $f(h)$. For the current frequency estimation problem, the right-hand side of Equation (35) can be computed, with some simple manipulations, as

$$\mathbf{J}_F^{-1} \cdot \Gamma_{1.5}(0.5 \cdot K \cdot SNR) + \sigma_\theta^2 \cdot 2\Phi(\sqrt{K \cdot SNR}) \tag{36}$$

where $\mathbf{J}_F$ is the Fisher information matrix computed as (equation 82, Appendix A) and $\sigma_\theta^2$ is variance related to the *a priori* distribution of $\theta$. $\Phi(z)$ is the tail probability of the standard normal distribution at point $z$. and $\Gamma_a(z)$ is the incomplete gamma function defined by

$$\Gamma_a(z) = \frac{1}{\Gamma(a)}\int_0^z e^{-v} v^{a-1} dv \tag{37}$$

herein. Comparison between WWB and ZZB are shown in FIGURE 12. The WWB configuration is again the proposed $\text{WWB}_{(2,9,10)}$. In all cases WWB and ZZB converge both asymptotically (in high SNR) and in the no-information region.

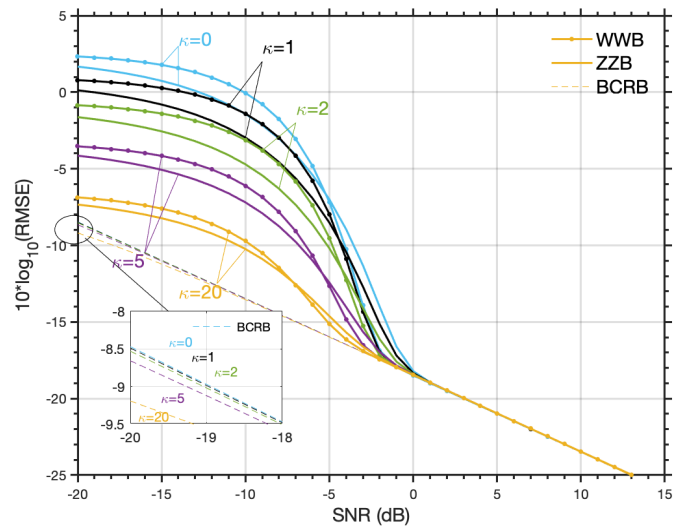

**FIGURE 12** WWB and ZZB with von Mises distribution as *a priori* pdf: concentration parameter $\kappa \in \{0,1,2,5,20\}$ and mean $\mu = 0$.

In FIGURE 13, we compare both WWB and ZZB to MAP. Both of them are tight bounds in the sense that the difference between them and MAP is small within the Barankin and threshold regions.

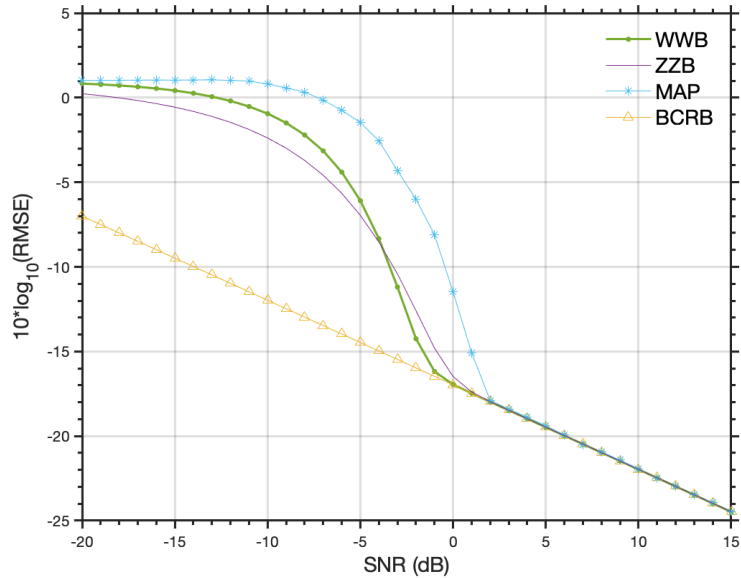

**FIGURE 13** WWB vs ZZB. $\kappa = 1$, $\mu = 0$. RMSE for MAP estimates are also plotted as performance benchmarks.

One interesting observation emerges. ZZB tend to be larger (thus tighter) than the WWB very close to the threshold, but smaller (thus looser) than the developed WWB for lower values of SNR. The larger the value of $\kappa$, the less this difference is (FIGURE 13). Existing work tends to discuss threshold behavior of bounds in a spanned region hence we divide the threshold region (e.g. qualitatively defined in Weiss's 4-region model mentioned in FIGURE 1) into two sub-regions:

Region 1 (SNR = -10 ~ -3.5 dB): WWB > ZZB. WWB outperforms ZZB. The largest performance gain occurs at -8 dB (C/N0=22 dB-Hz) and amounts to 1.5 dB. Take GPS L1 C/A for example, this means a 67 Hz improvement in bounding accuracy: $f_{INT} \cdot |RMSE_{WWB} - RMSE_{ZZB}|/(2\pi)$, where $f_{INT} = 1000$ Hz. This is not a trivial improvement ($22.5\% = 10^{\frac{1.5}{10}}/(2\pi)$) in bounding accuracy (ZZB: 28.8 Hz, WWB: 95.7 Hz). Note also here that $RMSE_i (i = WWB, ZZB)$ represents the RMSE's of the two compared bounds and should be converted from dB to radians first.

Region 2 (SNR = -3.5 ~ 0 dB): WWB < ZZB. ZZB outperforms WWB with smaller performance gains compared to the previous case.

It is true that ZZB outperforms WWB in terms of predicting the exact thresholding *point* where outliers surge triggered by deteriorating SNR; this is a rather interesting observation. Meanwhile it is also true that within the majority of the threshold *region*, WWB succeeds in providing tighter bounds than ZZB. This alternating performances of the two most-discussed

Bayesian bounds are repeatedly confirmed in FIGURE 12 where the same prior pdf with different parameters is used.

The phenomenon means tradeoff has to be made between these two Bayesian type bounds. For extremely low SNR (from -3.5 dB down to -20 dB), WWB is a better performance indicator, while in a relatively limited range of SNR (between -3.5~0 dB), ZZB is preferred.

## 6. CONCLUSIONS

In this work, we developed a Weiss-Weinstein bound (WWB) for circular frequency tracking error in GNSS receivers, by deriving the bound and proposing a new method to select a new set of test points that is required by WWB. We choose WWB in preference over other modern Bayesian bounds for its being free from regularity conditions or requirements on prior distribution. Later in the evaluation part we conclude that this divide-and-conquer type of selecting test points help to tighten the bound in the mid-to-low SNR region (-20 ~ -3.5 dB).

We proposed to use von Mises distribution, a cyclic counterpart of Gaussian distribution as prior distribution to specifically deal with cyclic nature of the estimated variable, i.e. carrier frequency in GNSS signals. This helps to tighten the bound in 'no signal' zone where signals are extremely weak (close to SNR=-20 dB).

We evaluated our methods through Monte Carlo simulations and compare it to Ziv- Zakaï bound and confirm that the developed bound can be tighter than Ziv-Zakai bound by up to 22.5% when SNR varies between –3.5 dB and -20 dB.

Future efforts may be invested in deriving the WWB for joint frequency and phase estimation and using the bound as a constraint condition in developing innovative code and carrier estimators (Pany, 2010) in weak GNSS conditions.

## ACKNOWLEDGEMENTS
This work was supported by the National Key R&D Program of China (2022YFB3904401).

**Conflict of Interests**
The authors declare no potential conflicts of interest.

**APPENDIX A. WWB**

The $\mathbf{Q}$ matrix of WWB was provided Equation (15) and is re-written here for ease of reference:

$$[\mathbf{Q}]_{ij} = \frac{\left[ E_{\mathbf{x},\theta}\left\{\frac{p^{s_i}(\mathbf{x},\theta+h_i)p^{s_j}(\mathbf{x},\theta+h_j)}{p^{s_i+s_j}(\mathbf{x},\theta)}\right\} - E_{\mathbf{x},\theta}\left\{\frac{p^{s_i-s_j-1}(\mathbf{x},\theta)p^{s_j}(\mathbf{x},\theta+h_j)}{p^{s_i-1}(\mathbf{x},\theta-h_i)}\right\} \right.}{E_{\mathbf{x},\theta}\left\{\frac{p^{s_i}(\mathbf{x},\theta+h_i)}{p^{s_i}(\mathbf{x},\theta)}\right\} \cdot E_{\mathbf{x},\theta}\left\{\frac{p^{s_j}(\mathbf{x},\theta+h_j)}{p^{s_j}(\mathbf{x},\theta)}\right\}} \left. \frac{-E_{\mathbf{x},\theta}\left\{\frac{p^{s_j-s_i-1}(\mathbf{x},\theta)p^{s_i}(\mathbf{x},\theta+h_i)}{p^{s_j-1}(\mathbf{x},\theta-h_j)}\right\} + E_{\mathbf{x},\theta}\left\{\frac{p^{s_i+s_j-2}(\mathbf{x},\theta)}{p^{s_i-1}(\mathbf{x},\theta-h_i)p^{s_j-1}(\mathbf{x},\theta-h_j)}\right\}}{} \right] \quad (38)$$

We will evaluate each expectation in the denominator and nominator separately, using von Mises as *a prior* distribution.

**A.1 The denominator of Equation (38)**

The denominator is

$$E_{\mathbf{x},\theta}\{L^{s_i}(\mathbf{x};\theta+h_i,\theta)\}E_{\mathbf{x},\theta}\{L^{s_j}(\mathbf{x};\theta+h_j,\theta)\} \\ = E_{\mathbf{x},\theta}\{L^{s_i}(\mathbf{x};\theta+h_i,\theta)\}E_{\mathbf{x},\theta}\{L^{s_j}(\mathbf{x};\theta+h_j,\theta)\} \quad (39)$$

By inspection, the denominator is a product of two expectations with identical likelihood ratio structures; only the test points ($h_i$ and $h_j$) and thus the subscripts differ. We shall derive the first expectation and obtain the second expectation with a change of subscripts. The first part of the product can be evaluated using a double integration as:

$$\begin{aligned}E_{\mathbf{x},\theta}\{L^{s_i}(\mathbf{x};\theta+h_i,\theta)\} &= E_{\mathbf{x},\theta}\left\{\frac{p^{s_i}(\mathbf{x},\theta+h_i)}{p^{s_i}(\mathbf{x},\theta)}\right\} \\ &= \int_{\Omega_\theta}\int_{\Omega_\mathbf{x}}\frac{p^{s_i}(\mathbf{x},\theta+h_i)}{p^{s_i}(\mathbf{x},\theta)}p(\mathbf{x},\theta)d\mathbf{x}d\theta \\ &= \int_{\Omega_\theta}p^{1-s_i}(\theta)p^{s_i}(\theta+h_i)[\int_{\Omega_\mathbf{x}}p^{1-s_i}(\mathbf{x}\mid\theta)p^{s_i}(\mathbf{x}\mid\theta+h_i)d\mathbf{x}]d\theta\end{aligned} \quad (40)$$

where $\Omega_\theta$ and $\Omega_\mathbf{x}$ are all possible values of $\theta$ and $\mathbf{x}$, respectively. By the end of A.1.1, we will prove that this double integral can be in fact evaluated using two independent integrals. The rest of derivations in this appendix will benefit greatly from this fact.

### A.1.1 The inner integral

The inner integral of equation (40) can be evaluated as

$$\begin{aligned}&\int_{\Omega_\mathbf{x}}p^{1-s_i}(\mathbf{x}\mid\theta)p^{s_i}(\mathbf{x}\mid\theta+h_i)d\mathbf{x} \\ &= \int_{\Omega_\mathbf{x}}\frac{(2\pi\sigma^2)^{-Ks_i}e^{-\frac{s_i}{2\sigma^2}\sum_{k=0}^{K-1}\left(\|x_k\|^2+\|A\|^2-2\text{Re}\left\{A^*x_k e^{-j((\theta+h_i)k+\varphi)}\right\}\right)}}{(2\pi\sigma^2)^{-K(s_i-1)}e^{-\frac{s_i-1}{2\sigma^2}\sum_{k=0}^{K-1}\|x_k\|^2+\|A\|^2-2\text{Re}\left\{A^*x_k e^{-j(\theta\cdot k+\varphi)}\right\}}}d\mathbf{x} \\ &= \int_{\Omega_\mathbf{x}}(2\pi\sigma^2)^{-K}e^{-\frac{1}{2\sigma^2}\sum_{k=0}^{K-1}(\|x_k\|^2+\|A\|^2-2\text{Re}\left\{A^*e^{-j(\theta\cdot k+\varphi)}x_k[s_i e^{-jh_i k}-(s_i-1)]\right\})}d\mathbf{x}\end{aligned} \quad (41)$$

Let us set

$$y_k = e^{-j\varphi}\cdot x_k - A\cdot e^{j\theta\cdot k}\cdot a_k \quad (42)$$

where $a_k$ is set as $s_i(e^{jh_i k}-1)+1$, which is a deterministic part (remember we are deriving formulas of probability conditioned on the frequency to be estimated) and $y_k$ is thus a complex (circular) Gaussian variable due to the circular symmetric and Gaussian assumption on $x_k$. Hence,

$$\begin{aligned}\|a_k\|^2 &= [s_i(e^{jh_i k}-1)+1][s_i(e^{-jh_i k}-1)+1] \\ &= 2s_i(1-s_i)[\cos(h_i k)-1]+1,\end{aligned} \quad (43)$$

and

$$\begin{aligned}\|y_k\|^2 &= \|x_k\|^2 + \|A\|^2\|a_k\|^2 - 2\text{Re}\{e^{-j(\theta\cdot k+\varphi)}A^*x_k a_k^*\} \\ &= \|x_k\|^2 + \|A\|^2 + 2s_i(1-s_i)[\cos(h_i k)-1]\|A\|^2 \\ &\quad - 2\Re\{e^{-j(\theta\cdot k+\varphi)}A^*x_k[s_i e^{-jh_i k}-(s_i-1)]\}\end{aligned} \quad (44)$$

Substituting this back into Equation (41), we arrive at

$$\begin{aligned}&\int_{\Omega_\mathbf{x}}p^{1-s_i}(\mathbf{x}\mid\theta)p^{s_i}(\mathbf{x}\mid\theta+h_i)d\mathbf{x} \\ &= \int_{\Omega_\mathbf{x}}(2\pi\sigma^2)^{-K}e^{-\frac{1}{2\sigma^2}\sum_{k=0}^{K-1}\left(\|x_k\|^2+\|A\|^2-2\text{Re}\left\{A^*e^{-j(\theta\cdot k+\varphi)}x_k[s_i e^{-jh_i k}-(s_i-1)]\right\}\right)}d\mathbf{x} \\ &= \int_{\Omega_\mathbf{x}}(2\pi\sigma^2)^{-K}e^{-\frac{1}{2\sigma^2}\sum_{k=0}^{K-1}\left(\|y_k\|^2-2s_i(1-s_i)[\cos(h_i k)-1]\|A\|^2\right)}d\mathbf{x} \\ &= (2\pi\sigma^2)^{-K}e^{-\frac{1}{2\sigma^2}\sum_{k=0}^{K-1}(-2s_i(1-s_i)[\cos(h_i k)-1]\|A\|^2)}\int_{\Omega_\mathbf{x}}e^{-\frac{1}{2\sigma^2}\sum_{k=0}^{K-1}\|y_k\|^2}d\mathbf{x}\end{aligned} \quad (45)$$

where

$$\int_{\Omega_{\mathbf{x}}} e^{-\frac{1}{2\sigma^2}\sum_{k=0}^{K-1}\|y_k\|^2} d\mathbf{x} = \prod_{k=0}^{K-1}\int_{-\infty}^{+\infty} e^{-\frac{1}{2\sigma^2}\|e^{-j\varphi}x_k - Ae^{j\theta\cdot k}a_k\|^2} dx_k$$
$$= \prod_{k=0}^{K-1}\int_{-\infty}^{+\infty} e^{-\frac{1}{2\sigma^2}\|y_k\|^2} dy_k \qquad (46)$$
$$= \prod_{k=0}^{K-1}\int_{-\infty}^{+\infty} e^{-\frac{[\Re e\{y_k\}]^2 + [\Im m\{y_k\}]^2}{2\sigma^2}} dy_k$$
$$= (2\pi\sigma^2)^K,$$

where $\Re e\{\cdot\}$ and $\Im m\{\cdot\}$ represents real and imaginary part of a complex number. Substitute into Equation (45) and take the natural logarithm and recognizing $SNR = \frac{\|A\|^2}{2\sigma^2}$, we have

$$\ln \int_{\Omega_{\mathbf{x}}} p^{1-s_i}(\mathbf{x}\mid\theta) p^{s_i}(\mathbf{x}\mid\theta + h_i) d\mathbf{x} = -\frac{\|A\|^2}{2\sigma^2}\sum_{k=0}^{K-1}(-2s_i(1-s_i)[\cos(h_i k) - 1])$$
$$= -s_i(1-s_i)\cdot 2K \cdot SNR\left[1 - \frac{1}{K}\sum_{k=0}^{K-1}\cos(h_i k)\right] \qquad (47)$$

The summation in the square brackets can be evaluated as:

$$\sum_{k=0}^{K-1}\cos(h_i k) = \sum_{k=0}^{K-1}\Re e\{e^{-jh_i k}\} = \Re e\left\{\sum_{k=0}^{K-1} e^{-jh_i k}\right\} = \Re e\left\{\frac{1 - e^{-jh_i K}}{1 - e^{-jh_i}}\right\}$$
$$= \Re e\left\{e^{-j\frac{h_i(K-1)}{2}}\frac{\sin\left(\frac{h_i K}{2}\right)}{\sin\left(\frac{h_i}{2}\right)}\right\} = \cos\frac{h_i(K-1)}{2}\cdot\frac{\sin\left(\frac{h_i K}{2}\right)}{\sin\left(\frac{h_i}{2}\right)} \qquad (48)$$

This result allows for a handy notation of the internal integral:

$$\int_{\Omega_{\mathbf{x}}} p^{1-s_i}(\mathbf{x}\mid\theta) p^{s_i}(\mathbf{x}\mid\theta + h_i) d\mathbf{x} = e^{\mu_i} \qquad (49)$$

where $e^{\mu_i}$ is the Euler's number raised to the power of $\mu_i$ and

$$\mu_i = -s_i(1-s_i)\cdot 2K \cdot SNR\left(1 - \frac{1}{K}\cos\left[\frac{h_i(K-1)}{2}\right]\cdot\frac{\sin\left[\frac{h_i K}{2}\right]}{\sin\left[\frac{h_i}{2}\right]}\right) \qquad (50)$$

With this identity, it turns out that Equation (40), which is a double integration, can be split into two separate integrations: one of them equals exactly $\int_{\Omega_{\mathbf{x}}} p^{1-s_i}(\mathbf{x}|\theta) p^{s_i}(\mathbf{x}|\theta + h_i) d\mathbf{x}$, and the other is $\int_{\Omega_\theta} p^{1-s_i}(\theta) p^{s_i}(\theta + h_i) d\theta$. The latter will be evaluated in A.1.2.

### A.1.2 The outer integral

The outer integral of the double integration can be evaluated using the a priori information about $f$, as:

$$\int_{\Omega_\theta} p^{1-s_i}(\theta) p^{s_i}(\theta + h_i) d\theta = \int_{-\pi}^{\pi - h_i}\frac{1}{2\pi I_0(\kappa)} e^{\kappa[(1-s_i)\cdot\cos(\theta-\mu) + s_i\cdot\cos(\theta+h_i-\mu)]} d\theta \qquad (51)$$

To comply with the notation adopted for the result of the first integral, we denote $\ln\int_{-\pi}^{\pi-h_i}\frac{1}{2\pi I_0(\kappa)} e^{[(1-s_i)\cdot\cos(\theta-\mu) + s_i\cdot\cos(\theta+h_i-\mu)]}$ as $\gamma_i$:

$$\gamma_i = \ln\int_{-\pi}^{\pi-h_i}\frac{1}{2\pi I_0(\kappa)} e^{\kappa[(1-s_i)\cdot\cos(\theta-\mu) + s_i\cdot\cos(\theta+h_i-\mu)]} d\theta$$
$$= \ln\int_0^{1-h}\frac{1}{I_0(\kappa)} e^{\kappa[(s_i-1)\cdot\cos(2\pi t-\mu) - s_i\cdot\cos(2\pi t-\mu+h_i)]} dt \qquad (52)$$

where the second equality applies changes of variable, $t = \frac{\theta+\pi}{2\pi}$ and $h = \frac{h_i}{2\pi}$. Then the second integral is $e^{\gamma_i}$ and the left hand side of Equation (40) can finally be evaluated as $e^{\mu_i}e^{\gamma_i}$. With a simple change of subscript from $i$ to $j$, $\mu_j$ and $\gamma_j$ can be readily obtained. The denominator of $[\mathbf{Q}]_{ij}$ can thus be evaluated as

$$E_{\mathbf{x},\theta}\{L^{s_i}(\mathbf{x};\theta+\mathbf{h}_i,\theta)\}E_{\mathbf{x},\theta}\{L^{s_j}(\mathbf{x};\theta+\mathbf{h}_j,\theta)\} = e^{\mu_i+\mu_j}e^{\gamma_i+\gamma_j} \tag{53}$$

The integration and $I_0(\kappa)$ necessary to compute $\gamma_i$ and $\gamma_j$ are easily done with available Matlab built-in functions.

### A.2 The nominator of Equation (38)

The nominator of (38) becomes

$$\begin{aligned} &E_{\mathbf{x},\theta}\left\{\frac{p^{s_i}(\mathbf{x},\theta+h_i)p^{s_j}(\mathbf{x},\theta+h_j)}{p^{s_i+s_j}(\mathbf{x},\theta)}\right\} - E_{\mathbf{x},\theta}\left\{\frac{p^{s_j}(\mathbf{x},\theta+h_j)}{p^{1-s_i+s_j}(\mathbf{x},\theta)p^{s_i-1}(\mathbf{x},\theta-h_i)}\right\} \\ &-E_{\mathbf{x},\theta}\left\{\frac{p^{s_i}(\mathbf{x},\theta+h_i)}{p^{1-s_j+s_i}(\mathbf{x},\theta)p^{s_j-1}(\mathbf{x},\theta-h_j)}\right\} + E_{\mathbf{x},\theta}\left\{\frac{p^{s_i+s_j-2}(\mathbf{x},\theta)}{p^{s_i-1}(\mathbf{x},\theta-h_i)p^{s_j-1}(\mathbf{x},\theta-h_j)}\right\} \end{aligned} \tag{54}$$

which is an algebraic sum of four expectations with respect to $p(\mathbf{x},\theta)$. Development of these four into their computation-friendly forms are similar. In the sequel we will derive the final form for the fourth expectation in detail while providing a brief treatment of the other three.

### A. 2.1 The 4$^{th}$ expectation from Equation (54)

We have

$$\begin{aligned} &E_{\mathbf{x},\theta}\left\{\frac{p^{s_i+s_j-2}(\mathbf{x},\theta)}{p^{s_i-1}(\mathbf{x},\theta-h_i)p^{s_j-1}(\mathbf{x},\theta-h_j)}\right\} \\ &= \int_{\Omega_\theta}\int_{\Omega_\mathbf{x}} \frac{p^{s_i+s_j-1}(\mathbf{x},\theta)}{p^{s_i-1}(\mathbf{x},\theta-h_i)p^{s_j-1}(\mathbf{x},\theta-h_j)} d\mathbf{x}d\theta \\ &= \int_{\Omega_\theta} p^{s_i+s_j-1}(\theta)p^{1-s_i}(\theta-h_i)p^{1-s_j}(\theta-h_j) \cdot \\ &\quad \left[\int_{\Omega_\mathbf{x}} p^{s_i+s_j-1}(\mathbf{x}\mid\theta)p^{1-s_i}(\mathbf{x}\mid(\theta-h_i))p^{1-s_j}(\mathbf{x}\mid(\theta-h_j)) d\mathbf{x}\right] d\theta \end{aligned} \tag{55}$$

Again, this is a double integration and can eventually be evaluated by a product of two separate integrals. We will prove this assertion by evaluating the inner and outer integrals, just like what we did in A.1.

$$\begin{aligned} &p^{s_i+s_j-1}(\mathbf{x}\mid\theta)p^{1-s_i}(\mathbf{x}\mid(\theta-h_i))p^{1-s_j}\left(\mathbf{x}\mid(\theta-h_j)\right) \\ &= (2\pi\sigma^2)^{-K}\exp\left\{-\frac{1}{2\sigma^2}\sum_{k=0}^{K-1}\left(\|x_k\|^2 + \|A\|^2\right.\right. \\ &\quad \left.\left. - 2\Re e\{A^*x_k e^{-j(\theta\cdot k+\varphi)}[(s_i+s_j-1)-(s_i-1)e^{jh_ik}-(s_j-1)e^{jh_jk}]\}\right)\right\} \end{aligned} \tag{56}$$

Just as we have already done in A.1.1, we set

$$y_k = e^{-j\varphi} \cdot x_k - A \cdot e^{j\theta\cdot k} \cdot a_k \tag{57}$$

where this time $a_k$ is defined as

$$a_k^* = (s_i+s_j-1)e^{jh_ik} - (s_i-1)e^{jh_ik} - (s_j-1)e^{jh_jk} \tag{58}$$

to facilitate derivation. Thus,

$$\begin{aligned}\|a_k\|^2 = {}& (s_i+s_j-1)^2+(s_i-1)^2+(s_j-1)^2\\&-2\cdot(s_i+s_j-1)\cdot(s_i-1)\cdot\cos(h_ik)\\&-2\cdot(s_i+s_j-1)\cdot(s_j-1)\cdot\cos(h_jk)\\&+2\cdot(s_i-1)\cdot(s_j-1)\cdot\cos[(h_i-h_j)k]-1+1.\end{aligned} \qquad (59)$$

This leads to

$$\begin{aligned}\|y_k\|^2 &= \|x_k\|^2+\|A\|^2\|a_k\|^2-2\Re e\{A^*x_ke^{-j(\theta\cdot k+\varphi)}a_k^*\}\\&= \|x_k\|^2+\|A\|^2-2\Re e\{A^*x_ke^{-j(\theta\cdot k+\varphi)}a_k^*\}\\&\quad +\|A\|^2\cdot D_k\end{aligned} \qquad (60)$$

where $D_k$ is a quantity dependent on $k$ as

$$\begin{aligned}D_k = \big[&(s_i+s_j-1)^2+(s_i-1)^2+(s_j-1)^2-1\\&-2\cdot(s_i+s_j-1)\cdot(s_i-1)\cdot\cos(h_ik)\\&-2\cdot(s_i+s_j-1)\cdot(s_j-1)\cdot\cos(h_jk)\\&+2\cdot(s_i-1)\cdot(s_j-1)\cdot\cos[(h_i-h_j)k]\big]\end{aligned} \qquad (61)$$

Substitute back into Equation (56) and we arrive at:

$$\begin{aligned}&\int_{\Omega_{\mathbf{x}}}p^{s_i+s_j-1}(\mathbf{x}\mid\theta)p^{1-s_i}(\mathbf{x}\mid\theta-h_i)p^{1-s_j}(\mathbf{x}\mid\theta-h_j)d\mathbf{x}\\&=\int_{\Omega_{\mathbf{x}}}(2\pi\sigma^2)^{-K}e^{-\frac{1}{2\sigma^2}\sum_{k=0}^{K-1}\left(\|x_k\|^2+\|A\|^2\|a_k\|^2-2\mathrm{Re}\{A^*x_ke^{-j(\theta\cdot k+\varphi)}a_k^*\}\right)}d\mathbf{x}\\&=\int_{\Omega_{\mathbf{x}}}(2\pi\sigma^2)^{-K}e^{-\frac{1}{2\sigma^2}\sum_{k=0}^{K-1}\left(\|y_k\|^2-\|A\|^2\cdot D_k\right)}d\mathbf{x}\\&=(2\pi\sigma^2)^{-K}e^{\frac{1}{2\sigma^2}\sum_{k=0}^{K-1}D_k\|A\|^2}\int_{\mathbf{x}}e^{-\frac{1}{2\sigma^2}\sum_{k=0}^{K-1}|y_k\|^2}d\mathbf{x}\\&=e^{\frac{\|A\|^2}{2\sigma^2}\sum_{k=0}^{K-1}D_k}.\end{aligned} \qquad (62)$$

The last equality is based on identity (46) and recognizing $SNR=\frac{\|A\|^2}{2\sigma^2}$. Take the natural logarithm of both sides of Equation (62), and we will arrive at

$$\int_{\Omega_{\mathbf{x}}}p^{s_i+s_j-1}(\mathbf{x}\mid\theta)p^{1-s_i}(\mathbf{x}\mid\theta-h_i)p^{1-s_j}(x\mid\theta-h_j)d\mathbf{x}=e^{\mu_{ij,4}} \qquad (63)$$

where $\mu_{ij,4}$ is defined as

$$\begin{aligned}\mu_{ij,4} = {}& SNR\cdot\Big[K\cdot\left((s_i+s_j-1)^2+(s_i-1)^2+(s_j-1)^2-1\right)\\&-2\cdot(s_i+s_j-1)\cdot(s_i-1)\cdot\frac{\cos\left[\frac{h_i(K-1)}{2}\right]\cdot\sin\left[\frac{h_iK}{2}\right]}{\sin\left[\frac{h_i}{2}\right]}\\&-2\cdot(s_i+s_j-1)\cdot(s_j-1)\cdot\frac{\cos\left[\frac{h_j(K-1)}{2}\right]\cdot\sin\left[\frac{h_jK}{2}\right]}{\sin\left[\frac{h_j}{2}\right]}\\&+2\cdot(s_i-1)\cdot(s_j-1)\cdot\frac{\cos\left[\frac{(h_i-h_j)(K-1)}{2}\right]\cdot\sin\left[\frac{(h_i-h_j)K}{2}\right]}{\sin\left[\frac{(h_i-h_j)}{2}\right]}.\end{aligned} \qquad (64)$$

Since $\mu_{ij,4}$ does not depend on $\theta$, Equation (55) can be computed as a product of two integrals, one of which takes the value of $e^{\mu_{ij,4}}$:

$$\int_{\Omega_\theta} p^{s_i+s_j-1}(\theta)p^{1-s_i}(\theta-h_i)p^{1-s_j}(\theta-h_j)\left[\int_{\Omega_x} p^{s_i+s_j-1}(\mathbf{x}\mid\theta)p^{1-s_i}(\mathbf{x}\mid(\theta-h_i))p^{1-s_j}(\mathbf{x}\mid(\theta-h_j))\,d\mathbf{x}\right]d\theta \quad (65)$$

$$= e^{\mu_{ij,4}}\int_{\Omega_\theta} p^{s_i+s_j-1}(\theta)p^{1-s_i}(\theta-h_i)p^{1-s_j}(\theta-h_j)\,d\theta$$

Now, the integral over $\Omega_\theta$ can be computed using the a priori pdf as

$$\begin{aligned}
&\int_{\Omega_\theta} p^{s_i+s_j-1}(\theta)p^{1-s_i}(\theta-h_i)p^{1-s_j}(\theta-h_j)\,d\theta\\
&= \int_{-\pi+h_i}^{\pi}\frac{1}{2\pi I_0(\kappa)}e^{\kappa[(s_i+s_j-1)\cdot\cos(\theta-\mu)+(1-s_i)\cdot\cos(\theta-h_i-\mu)+(1-s_j)\cdot\cos(\theta-h_j-\mu)]}\,d\theta\\
&= \int_{\frac{h_j}{2\pi}}^{1-h}\frac{1}{I_0(\kappa)}e^{\kappa[(1-s_i-s_j)\cdot\cos(2\pi(t+h)-\mu)+(s_i-1)\cdot\cos(2\pi(t+h)-\mu-h_i)+(s_j-1)\cdot\cos(2\pi(t+h)-\mu-h_j)]}\,dt\\
&= e^{\gamma_{ij,4}}
\end{aligned} \quad (66)$$

where, without loss of generality, we assume $h_i > h_j > 0$ and the second equality applies changes of variable, $h = \frac{h_i - h_j}{2\pi}$ and $t = \frac{\theta+\pi}{2\pi} - h$. Insert Equation (66) into (65), the fourth expectation of Equation (54) is computed as

$$E_{\mathbf{x},\theta}\left\{\frac{p^{s_i+s_j-2}(\mathbf{x},\theta)}{p^{s_i-1}(\mathbf{x},\theta-h_i)p^{s_j-1}(\mathbf{x},\theta-h_j)}\right\} = e^{\mu_{ij,4}}e^{\gamma_{ij,4}},\quad h_i > h_j > 0 \quad (67)$$

The first, second and third expectations can be evaluated in a similar manner.

### A. 2.2 The 1st expectation from Equation (54)

With similar manipulations, the first expectation in Equation (54) can be evaluated as

$$E_{\mathbf{x},\theta}\left\{\frac{p^{s_i}(\mathbf{x},\theta+h_i)p^{s_j}(\mathbf{x},\theta+h_j)}{p^{s_i+s_j}(\mathbf{x},\theta)}\right\} = e^{\mu_{ij,1}}e^{\gamma_{ij,1}} \quad (68)$$

where $\gamma_{ij,1}$ and $\mu_{ij,1}$ is defined as:

$$\begin{aligned}
&\int_{\Omega_\theta} p^{1-s_i-s_j}(\theta)p^{s_i}(\theta+h_i)p^{s_j}(\theta+h_j)\,d\theta\\
&= \int_{-\pi}^{\pi-h_i}\frac{1}{2\pi I_0(\kappa)}e^{\kappa[(1-s_i-s_j)\cdot\cos(\theta-\mu)+s_i\cdot\cos(\theta+h_i-\mu)+s_j\cdot\cos(\theta+h_j-\mu)]}\,d\theta\\
&= \int_{\frac{h_j}{2\pi}}^{1-h}\frac{1}{I_0(\kappa)}e^{\kappa[(s_i+s_j-1)\cdot\cos(2\pi t-\mu-h_j)-s_i\cdot\cos(2\pi(t+h)-\mu)-s_j\cdot\cos(2\pi t-\mu)]}\,dt\\
&= e^{\gamma_{ij,1}}
\end{aligned} \quad (69)$$

and

$$\ln \int_{\Omega_\mathbf{x}} p^{1-s_i-s_j}(\mathbf{x} \mid \theta) p^{s_i}(\mathbf{x} \mid \theta + h_i) p^{s_j}(x \mid \theta + h_j) d\mathbf{x}$$

$$= SNR \cdot \Big[ K \cdot \big((s_i + s_j - 1)^2 + s_i^2 + s_j^2 - 1\big)$$

$$+ 2 \cdot s_i \cdot s_j \cdot \frac{\cos\left[\frac{(h_i - h_j)(K-1)}{2}\right] \cdot \sin\left[\frac{(h_i - h_j)K}{2}\right]}{\sin\left[\frac{(h_i - h_j)}{2}\right]}$$

$$- 2 \cdot (s_i + s_j - 1) \cdot s_i \cdot \frac{\cos\left[\frac{h_i(K-1)}{2}\right] \cdot \sin\left[\frac{h_i K}{2}\right]}{\sin\left[\frac{h_i}{2}\right]} \quad (70)$$

$$- 2 \cdot (s_i + s_j - 1) \cdot s_j \cdot \frac{\cos\left[\frac{h_j(K-1)}{2}\right] \cdot \sin\left[\frac{h_j K}{2}\right]}{\sin\left[\frac{h_j}{2}\right]}$$

$$= \mu_{ij,1}$$

respectively. During the development, $h_i > h_j > 0$ and the second equality of Equation (69) applies changes of variable, $h = \frac{h_i - h_j}{2\pi}$ and $t = \frac{\theta + \pi + h_j}{2\pi}$. Adoption of intermediate equality such as $a_k = s_i e^{jh_i k} + s_j e^{jh_j k} - (s_i + s_j - 1)$ and $y_k = e^{-j\varphi} x_k - A e^{j\theta \cdot k} a_k$ was helpful in arriving at Equation (70).

### A. 2.3 The 2$^{nd}$ expectation from Equation (54)

The second expectation from Equation (54) can be evaluated as

$$E_{\mathbf{x},\theta} \left\{ \frac{p^{s_i - s_j - 1}(\mathbf{x}, \theta) p^{s_j}(\mathbf{x}, \theta + h_j)}{p^{s_i - 1}(\mathbf{x}, \theta - h_i)} \right\} = e^{\mu_{ij,2}} e^{\gamma_{ij,2}} \quad (71)$$

where $e^{\gamma_{ij,2}}$ are defined as

$$\int_{\Omega_\theta} p^{s_i - s_j}(\theta) p^{s_j}(\theta + h_j) p^{1-s_i}(\theta - h_i) d\theta$$

$$= \int_{-\pi + h_i}^{\pi - h_j} \frac{1}{2\pi I_0(\kappa)} e^{\kappa[(s_i - s_j) \cdot \cos(\theta - \mu) + s_j \cdot \cos(\theta + h_j - \mu) + (1 - s_i) \cdot \cos(\theta - h_i - \mu)]} d\theta \quad (72)$$

With changes of variables $h = \frac{h_i + h_j}{2\pi}$ and $t = \frac{\theta + \pi - h_i}{2\pi}$, it can be shown

$$e^{\gamma_{ij,2}} = \int_0^{1-h} \frac{1}{I_0(\kappa)} e^{\kappa[(s_j - s_i) \cdot \cos(2\pi t + h_i - \mu) - s_j \cdot \cos(2\pi (t+h) - \mu) + (s_i - 1) \cdot \cos(2\pi t - \mu)]} dt \quad (73)$$

$\mu_{ij,2}$ of Equation (71) are defined as

$$\mu_{ij,2} = SNR \cdot \{ K \cdot \big(s_j^2 + (s_i - 1)^2 + (s_i - s_j)^2 - 1\big)$$

$$- 2 \cdot s_j \cdot (s_i - 1) \cdot \frac{\cos\left[\frac{(h_i + h_j)(K-1)}{2}\right] \cdot \sin\left[\frac{(h_i + h_j)K}{2}\right]}{\sin\left[\frac{(h_i + h_j)}{2}\right]}$$

$$+ 2 \cdot s_j \cdot (s_i - s_j) \cdot \frac{\cos\left[\frac{h_j(K-1)}{2}\right] \cdot \sin\left[\frac{h_j K}{2}\right]}{\sin\left[\frac{h_j}{2}\right]} \quad (74)$$

$$- 2 \cdot (s_i - 1) \cdot (s_i - s_j) \cdot \frac{\cos\left[\frac{h_i(K-1)}{2}\right] \cdot \sin\left[\frac{h_i K}{2}\right]}{\sin\left[\frac{h_i}{2}\right]} \}$$

During the development, the intermediate variable $y_k$ is set exactly as Equation (57), where $a_k$ is set as in

$$a_k^* = (s_i - s_j) + s_j e^{-jh_j k} - (s_i - 1)e^{jh_i k} \tag{75}$$

for arriving at Equation (74).

### A. 2.4 The 3$^{rd}$ expectations from Equation (54)

The third expectation from Equation (54) can be evaluated as

$$E_{\mathbf{x},\theta}\left\{\frac{p^{s_j - s_i - 1}(\mathbf{x}, \theta) p^{s_i}(\mathbf{x}, \theta + h_i)}{p^{s_j - 1}(\mathbf{x}, \theta - h_j)}\right\} = e^{\mu_{ij,3}} e^{\gamma_{ij,3}} \tag{76}$$

where $e^{\gamma_{ij,3}}$ and $\mu_{ij,3}$ can be evaluated easily by simply swapping the subscripts in Equations (73) and (74), respectively, as

$$e^{\gamma_{ij,3}} = \int_0^{1-h} \frac{1}{I_0(\kappa)} e^{\kappa[(s_i - s_j)\cdot\cos(2\pi t + h_j - \mu) - s_i \cdot \cos(2\pi(t+h) - \mu) + (s_j - 1)\cdot\cos(2\pi t - \mu)]} dt \tag{77}$$

and

$$\begin{aligned}
\mu_{ij,3} = \ &SNR \cdot \Bigg[K \cdot \left(s_i^2 + (s_j - 1)^2 + (s_i - s_j)^2 - 1\right) \\
&- 2 \cdot s_i \cdot (s_j - 1) \cdot \frac{\cos\left[\frac{(h_i + h_j)(K-1)}{2}\right] \cdot \sin\left[\frac{(h_i + h_j)K}{2}\right]}{\sin\left[\frac{(h_i + h_j)}{2}\right]} \\
&+ 2 \cdot s_i \cdot (s_j - s_i) \cdot \frac{\cos\left[\frac{h_i(K-1)}{2}\right] \cdot \sin\left[\frac{h_i K}{2}\right]}{\sin\left[\frac{h_i}{2}\right]} \\
&- 2 \cdot (s_j - 1) \cdot (s_j - s_i) \cdot \frac{\cos\left[\frac{h_j(K-1)}{2}\right] \cdot \sin\left[\frac{h_j K}{2}\right]}{\sin\left[\frac{h_j}{2}\right]}\Bigg]
\end{aligned} \tag{78}$$

This completes our development.

## APPENDIX B. BCRB

We can disassemble the $\mathbf{J}_B$ term of Equation (27) into two parts (van Trees & Bell, 2007) as

$$\mathbf{J}_B = \mathbf{J}_D + \mathbf{J}_P \tag{79}$$

Specifically, the subscript 'D' stands for 'data' and 'P' for 'a priori', which states that $\mathbf{J}_D$ represents the contribution from data and $\mathbf{J}_P$ the contribution from a priori pdf of the parameter to be estimated. They are defined as

$$\begin{aligned}
[\mathbf{J}_D]_{ij} &= E_{\mathbf{x},\boldsymbol{\theta}}\left\{\frac{\partial \ln p(\mathbf{x} \mid \boldsymbol{\theta})}{\partial \theta_i} \cdot \frac{\partial \ln p(\mathbf{x} \mid \boldsymbol{\theta})}{\partial \theta_j}\right\} \\
&= E_{\boldsymbol{\theta}}\left\{-E_{\mathbf{x}\mid\boldsymbol{\theta}}\left\{\frac{\partial^2 \ln p(\mathbf{x} \mid \boldsymbol{\theta})}{\partial \theta_i \, \partial \theta_j}\right\}\right\} \\
&= E_{\boldsymbol{\theta}}\{[\mathbf{J}_F]_{ij}\},
\end{aligned} \tag{80}$$

and

$$[\mathbf{J}_P]_{ij} = E_\mathbf{\theta} \left\{ \frac{\partial \ln p(\mathbf{\theta})}{\partial \theta_i} \cdot \frac{\partial \ln p(\mathbf{\theta})}{\partial \theta_j} \right\}$$
$$= -E \left\{ \frac{\partial^2 \ln p(\mathbf{\theta})}{\partial \theta_i \, \partial \theta_j} \right\} \tag{81}$$

from which we can readily identify the part bracketed by the outer curly braces of Equation (80) as the Fisher information matrix (FIM), denoted as $\mathbf{J}_F$. Assuming the first sample is taken at $t = 0$, the FIM for our current problem can be calculated as (Wu et el., 2006)

$$\mathbf{J}_F = SNR \frac{K(K-1)(2K-1)}{3} \tag{82}$$

leading to the determination of $\mathbf{J}_D$

$$\mathbf{J}_D = E_\theta\{\mathbf{J}_F\} = SNR \frac{K(K-1)(2K-1)}{3} \tag{83}$$

The $\mathbf{J}_P$ term can be computed as

$$E\left[ -\frac{\partial^2 \ln p(\theta)}{\partial \theta^2} \right] = \int_{-\pi}^{\pi} \kappa \cos(\theta - \mu) \cdot \frac{e^{\kappa \cos(\theta - \mu)}}{2\pi I_0(\kappa)} d\theta = \frac{\kappa}{I_0(\kappa)} \int_{-\pi}^{\pi} \cos(\theta - \mu) \cdot \frac{e^{\kappa \cos(\theta - \mu)}}{2\pi} d\theta$$
$$= \kappa \frac{I_1(\kappa)}{I_0(\kappa)} \tag{84}$$

The last equation mark is due to the fact that we recognize the righthand side of this equation includes the *modified Bessel function of the first kind and order 1* (Mardia, 1999, Appendix 1, eqn (A.1)). Substitute Equations (83) and (84) back into (79), we arrive at Equation (29) as BIM.